\documentclass[12pt]{iopart}

\usepackage[square, numbers, comma, sort&compress]{natbib}  
\usepackage{multirow}
\usepackage{ragged2e}
\usepackage[table,xcdraw]{xcolor}
\usepackage{booktabs}
\usepackage{pifont}
\usepackage[font=small, labelfont=bf]{caption}
\usepackage{subcaption}
\usepackage{siunitx}
\usepackage{adjustbox}
\usepackage{verbatim} 
\usepackage{amssymb}
\usepackage{geometry}
\usepackage{lscape}
\usepackage{url}

\newcommand{\newblock}{}

\makeatother

\begin{document}
\title{Sub-Scalp EEG for Sensorimotor Brain-Computer Interface}

\author{T B Mahoney$^{1}$, D B Grayden$^{1,2}$, S E John$^1$}

\address{$^1$Department of Biomedical Engineering, University of Melbourne, Victoria, 3010, Australia.}
\address{$^2$ Graeme Clark Institute, University of Melbourne, Victoria 3010, Australia.}
\ead{tmmah@unimelb.edu.au}

\providecommand{\keywords}[1]{\textbf{\textit{Index terms---}} #1}

\begin{abstract} \\
\textbf{Objective:}  To establish sub-scalp electroencephalography (EEG) as a viable option for brain-computer interface (BCI) applications, particularly for chronic use, by demonstrating its effectiveness in recording and classifying sensorimotor neural activity.\\
\textbf{Approach:} Two experiments were conducted in this study. The first aim was to demonstrate the high spatial resolution of sub-scalp EEG through analysis of somatosensory evoked potentials in sheep models. The second focused on the practical application of sub-scalp EEG, classifying motor execution using data collected during a sheep behavioural experiment.  \\
\textbf{Main Results:}  We successfully demonstrated the recording of sensorimotor rhythms using sub-scalp EEG in sheep models. Important spatial, temporal, and spectral features of these signals were identified, and we were able to classify motor execution with above-chance performance. These results are comparable to previous work that investigated signal quality and motor execution classification using ECoG and endovascular arrays in sheep models. \\
\textbf{Significance:}  These results suggest that sub-scalp EEG may provide signal quality that approaches that of more invasive neural recording methods such as ECoG and endovascular arrays, and support the use of sub-scalp EEG for chronic BCI applications.\\
\end{abstract}
\\

\maketitle

\section{Introduction}
Tetraplegia, or paralysis of the arms and legs, can result from stroke, brain/spinal cord trauma, and a range of neurodegenerative disorders, which are becoming increasingly common with the ageing global population. \citep{zedde_acute_2022, hacke_tetraplegia_1994}. For individuals with these disabilities, communication, mobility, and general interaction with their environment can be difficult. Assistive devices, such as joysticks and eye-trackers, use residual muscle control to allow the user to interface with wheelchairs, robotic prostheses, and computers \citep{sunny_eye-gaze_2021, bissoli_humanmachine_2019}. Rather than rely on limited muscle control, brain-computer interfaces (BCIs) can detect neural activity from paralysed muscles and use it to control devices, offering greater freedom of movement \citep{kubler_patients_2005, han_yuan_braincomputer_2014}.Despite the benefit these systems can provide, EEG BCIs are not used outside of research environments. 

Surveys of potential BCI users have established that lengthy donning and doffing procedures, maintenance, instability, and aesthetics of EEG systems make for poor chronic in-home use \citep{peters_brain-computer_2015, blabe_assessment_2015, kubler_user-centered_2014, kathner_multifunctional_2017, cardoso_comparing_2022}. Despite these pain points, users generally prefer non-invasive electrodes over implants \citep{brannigan_braincomputer_2024}. Endovascular stent-electrode arrays offer a minimally invasive alternative signal acquisition method for BCIs \citep{oxley_motor_2021}. However, currently endovascular BCIs cannot record activity from brain regions distal to large vasculature (namely the transverse and superior sagittal sinus), though ongoing research is focused on scaling stents to deploy in peripheral vessels \citep{brannigan_endovascular_2024, zhang_ultraflexible_2023}. Situating electrodes such that they record from independent neural sources is important for BCI functionality, as the user's modulation of these sources facilitates BCI control. Due to the poor spatial coverage of endovascular electrodes, current endovascular BCI are limited to a binary output, allowing the user to select output options when highlighted through an automated scrolling application or with eye-tracking-assisted selection. Additionally, the endovascular array cannot be removed post implantation in the event of failure, and carry risk of stenosis and thrombosis \cite{starke_endovascular_2015, modi_stent_2024, soldozy_systematic_2020}. There is a need for an alternative method of chronic EEG recording for BCI applications.

Sub-scalp EEG is a recent innovation in the neural signal acquisition space that addresses many of the limitations of traditional surface EEG setups and endovascular arrays. Sub-scalp EEG has already demonstrated safety and stability from chronic implantation in persons with epilepsy, and sufficient signal quality for seizure monitoring \citep{stirling_seizure_2021, duun-henriksen_new_2020, weisdorf_ultra-long-term_2019, barlatey_designing_2024, haneef_sub-scalp_2022}. Compared to external EEG caps, sub-scalp devices require no daily setup and are aesthetically discrete. Unlike endovascular arrays however, sub-scalp EEG can be removed and does not require access to critical vasculature. Additionally, sub-scalp arrays offer the flexibility to be positioned across the skull as needed, thereby accessing neural signals from cortical regions distal to major vasculature. This broader spatial coverage is anticipated to enable sub-scalp EEG to facilitate higher-dimensional BCI inputs, moving beyond the binary systems currently employed with endovascular arrays. Examples include two-dimensional \citep{dandan_huang_eeg-based_2009, kayagil_binary_2009} and three-dimensional cursor control \citep{mcfarland_electroencephalographic_2010} already demonstrated with non-invasive EEG, and may also extend to the rapidly developing field of speech decoding \citep{littlejohn_streaming_2025}. Previous studies have provided insights into sub-scalp EEG signal quality through comparison with alternative modalities. Sub-scalp EEG may record with similar signal quality, and thus BCI functionality, to non-invasive EEG, given the skull, rather than the scalp, is the major contributor of neural signal attenuation \citep{nunez_electric_2006}. Generally, when compared to non-invasive EEG recordings, sub-scalp EEG has demonstrated: similar or higher signal quality and reduced artefacts due to line noise, eye movements, and electrode shift, but similar or increased electromyography artefact, which is likely due to the proximity to the temporalis muscle \citep{young_comparison_2006, duun-henriksen_eeg_2015, weisdorf_high_2018, stirling_seizure_2021}. \citet{olson_comparison_2016} investigated similarities between sub-scalp EEG and ECoG, with a focus on signal bandwidth, demonstrating that sub-scalp EEG could record high-gamma activity up to 110~Hz, but with lower power than ECoG. Previous preliminary work has demonstrated comparable sub-scalp EEG signal amplitude, signal-to-noise ratio, and bandwidth to endovascular electrodes in sheep models \citep{mahoney_comparison_2023}. However, spatial resolution of sub-scalp EEG recordings is unclear. High spatial resolution is one of the primary motivators for intracranial recording methods such as electrocorticography (ECoG) over non-invasive EEG \citep{yang_high_2020, fifer_simultaneous_2014, fitzgerald_gamma_2018, trautner_sensory_2006, lachaux_many_2005, mainy_cortical_2008}. Characterising the spatial resolution of sub-scalp EEG recordings will provide insight into what performance to expect from a sub-scalp BCI device.

\begin{figure}[hbt]
\includegraphics[width=0.7\textwidth]{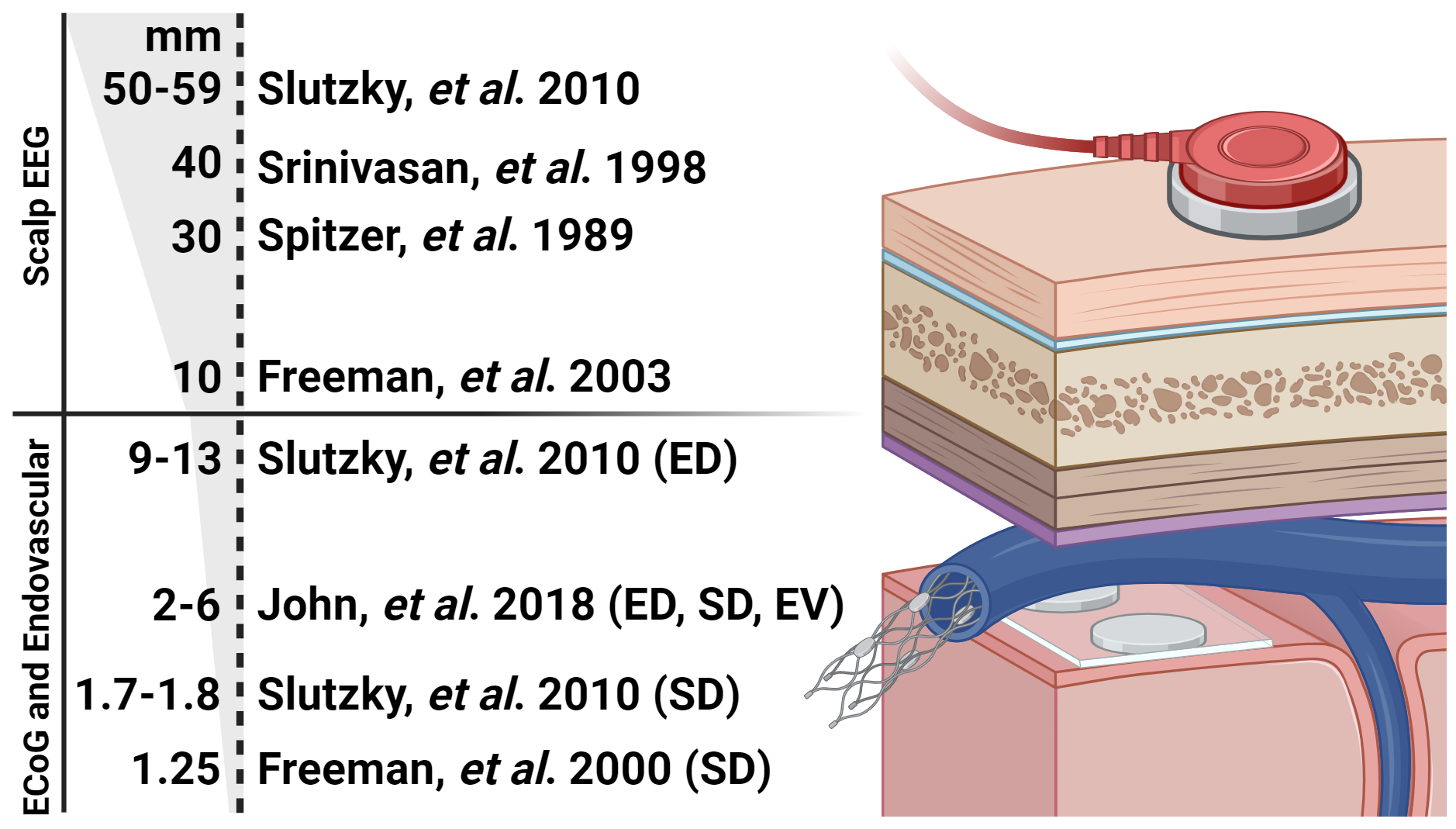}
\centering
\caption[Optimal Inter-electrode Distances in Literature]{ Outcomes of prior studies investigating optimal inter-electrode distances of neural recording methods: scalp EEG, ECoG (ED=epidural, SD=subdural), and endovascular (EV). (Created using biorender.com) }
\label{fig:CH5_Fig1}
\end{figure}

Previous work has explored spatial resolution with current signal acquisition methods (Figure~\ref{fig:CH5_Fig1}). Non-invasive EEG has typically demonstrated an optimal inter-electrode distance in the order of tens of millimetres \citep{slutzky_optimal_2010, robert_spitzer_method_1989, srinivasan_spatial_1998, freeman_spatial_2003}. Other work has investigated spatial resolutions of ECoG arrays \citep{slutzky_optimal_2010, john_signal_2018}, reporting optimal distances as low as 1.25~mm \citep{freeman_spatial_2000}.  More recent work has examined electrode size and spacing in novel endovascular electrode arrays and demonstrated equivalent spatial resolution with epidural and subdural ECoG \citep{john_signal_2018}. While we hypothesise that sub-scalp EEG will have a spatial resolution between non-invasive EEG and intracranial electrodes (1 to 10~mm), the optimal inter-electrode distance is unknown. 

This study comprises two aims toward investigating sub-scalp EEG feasibility for sensorimotor BCI applications. These are:
\begin{enumerate}
{\setlength\itemindent{40pt} \item [Aim 1:] Investigate the spatial resolution of sub-scalp EEG somatosensory evoked potentials (SEP).}
{\setlength\itemindent{40pt} \item [Aim 2:] Demonstrate sensorimotor activity decoding with sub-scalp EEG.}
\end{enumerate}

We performed \textit{in vivo} recording of SEPs in sheep models. SEPs are evoked in the somatosensory cortex, in response to  peripheral sensorimotor stimulus. In human models, the SEP waveform is often in the order of microvolts, and are generally localised to the posterior portion of the central sulcus \citep{schramm_localization_1991}. Being within a sulcus, SEP phases change depending on the relative location of the recording electrode. With sufficient signal quality, these phase changes can be used to locate the somatosensory cortex \citep{schreiner_mapping_2024}. The first aim of this study was to investigate the spatial resolution of sub-scalp EEG by examining spatial variations in SEP signal-to-noise ratio (SNR) and phase. 


The second aim was to demonstrate decoding of sensorimotor activity recorded from sub-scalp EEG in animal models to provide supporting evidence toward sBCI feasibility. We analysed sub-scalp EEG recordings from sheep models during a forced-choice behavioural task, and successfully decoded motor execution. We compare our results with an identical study performed previously with data recorded from an endovascular array \cite{forsyth_evaluation_2019} and report strong similarity in performance. Our research offers practical recommendations for the design of sBCI arrays, including guidelines for appropriate electrode size and spacing.  Furthermore, the successful decoding of sensorimotor activity confirms the feasibility of using sub-scalp EEG for brain-computer interface applications. Of the two aims, the findings of Aim 1 represent a significant advancement in our understanding of spatial resolution for electrodes placed in the sub-scalp space. While Aim 2 provides an interesting comparison to previous work with endovascular arrays, its findings are not as broadly generalizable as Aim 1, and its interpretation requires careful consideration of the experimental limitations, as described in Section \ref{limiations}. As such, greater detail and analysis were provided for Aim 1.

\section{Methods}
Both aims were investigated with adult female sheep models. The research was approved by the Animal Ethics Committee of the Florey Institute of Neuroscience and Mental Health, Melbourne, Australia (Approval number: 22010), which operates in accordance with the Australian Code for the care and use of animals for scientific purposes (8th edition 2013). Acute SEP stimulation experiments were performed to collect data for Aim 1. Aim 2 used data collected from sheep during a forced choice behavioural task. All analyses were performed in MATLAB (Version 2023a, MathWorks, USA).

\subsection{Acute SEP Experiment}
The purpose of this experiment was to investigate the spatial resolution of sub-scalp EEG SEP recordings. 

\subsubsection{Surgical Procedure}
During both the array implantation and the stimulation procedures, the sheep were induced using an intravenous injection of sodium thiopentone, and anaesthesia was maintained with isoflurane (1–2.0\% minimum alveolar concentration (MAC)).  The sheep were euthanised at the conclusion of each experiment via pentobarbitone injection. 

A 30~mm incision was made laterally across the scalp, posterior to the motor cortex. A spatula was used to separate the periosteum from the skull above the ovine sensorimotor cortix regions (as understood by previous research \citep{john_ovine_2017}) in an area large enough to allow insertion of the electrode arrays over the skull (approximately 30×30~mm). The arrays were placed laterally off centre by 10~mm, contralateral to the side of stimulation.

An incision was made in the forelimb midway between the elbow and knee joints. The electrode array used for stimulation (\Fref{fig:CH5_Fig3_a}) was inserted beneath the skin and wrapped around the musculature. After applying stimulation to each electrode individually, the electrode that most readily produced muscle contraction was used for the experiment. 

\subsubsection{Electrode Arrays}
The size of electrodes can impact spatial resolution. Small electrodes, intracranial electrodes in particular, can record signals of interest from local regions with little noise from more distal regions \citep{wolpaw_braincomputer_2012-2}. As electrode size increases, the electrodes record activity from regions of the brain that are outside of the region of interest, thereby increasing the noise component of the recording, and reducing spatial resolution. To account for this, we compare responses between arrays with different electrode diameters.

The electrode arrays were custom designs, manufactured on a two-layer flex PCB (polyimide, 0.22~mm thickness) by PCBway (China). Three array designs were used. The large array was a 3×3 grid with 10~mm inter-channel distance, or pitch, and 5~mm electrode diameter (\Fref{fig:CH5_Fig3_b}). The medium array was a 5×5 grid of electrodes with 5~mm pitch and 3~mm diameter (\Fref{fig:CH5_Fig3_c}). The small array was also a 5×5 grid with 5~mm, although with 1~mm electrode diameter (\Fref{fig:CH5_Fig3_d}). In this paper, arrays will be referred to by their electrode diameter. The arrays included ground and reference electrodes on the back (facing the scalp). All electrodes were gold plated at manufacture. Recording was performed using the G.USBAmp amplifier system (g.tec Medical Engineering, Austria), with a sampling frequency of 4800~Hz. The amplifier's high channel count and sampling frequency provided the necessary capabilities for this experiment.

\subsubsection{Stimulation Procedure}
\begin{figure}
    \centering
    \begin{subfigure}[b]{0.2\textwidth}
        \includegraphics[width=\textwidth]{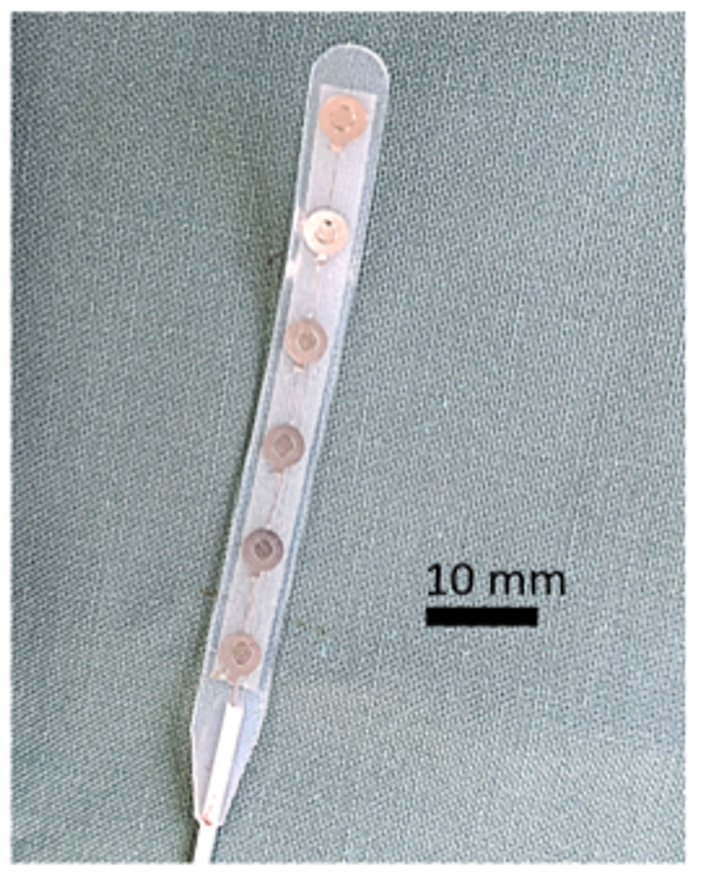}
        \caption{}
        \label{fig:CH5_Fig3_a}
    \end{subfigure} 
    \begin{subfigure}[b]{0.205\textwidth}
        \includegraphics[width=\textwidth]{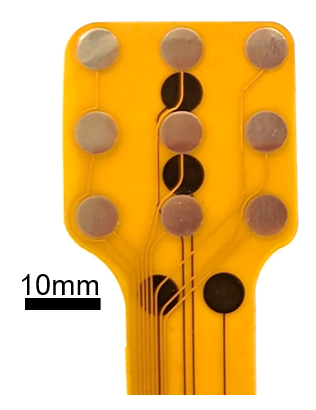}
        \caption{}
        \label{fig:CH5_Fig3_b}
    \end{subfigure}  
    \begin{subfigure}[b]{0.16\textwidth}
        \includegraphics[width=\textwidth]{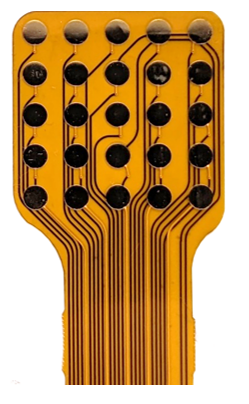}
        \caption{}
        \label{fig:CH5_Fig3_c}
    \end{subfigure} 
    \begin{subfigure}[b]{0.16\textwidth}
        \includegraphics[width=\textwidth]{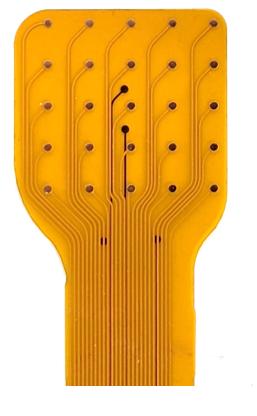}
        \caption{}
        \label{fig:CH5_Fig3_d}
    \end{subfigure} 
    
    \vspace{1mm}
    \begin{subfigure}[b]{0.76\textwidth}
        \includegraphics[width=\textwidth]{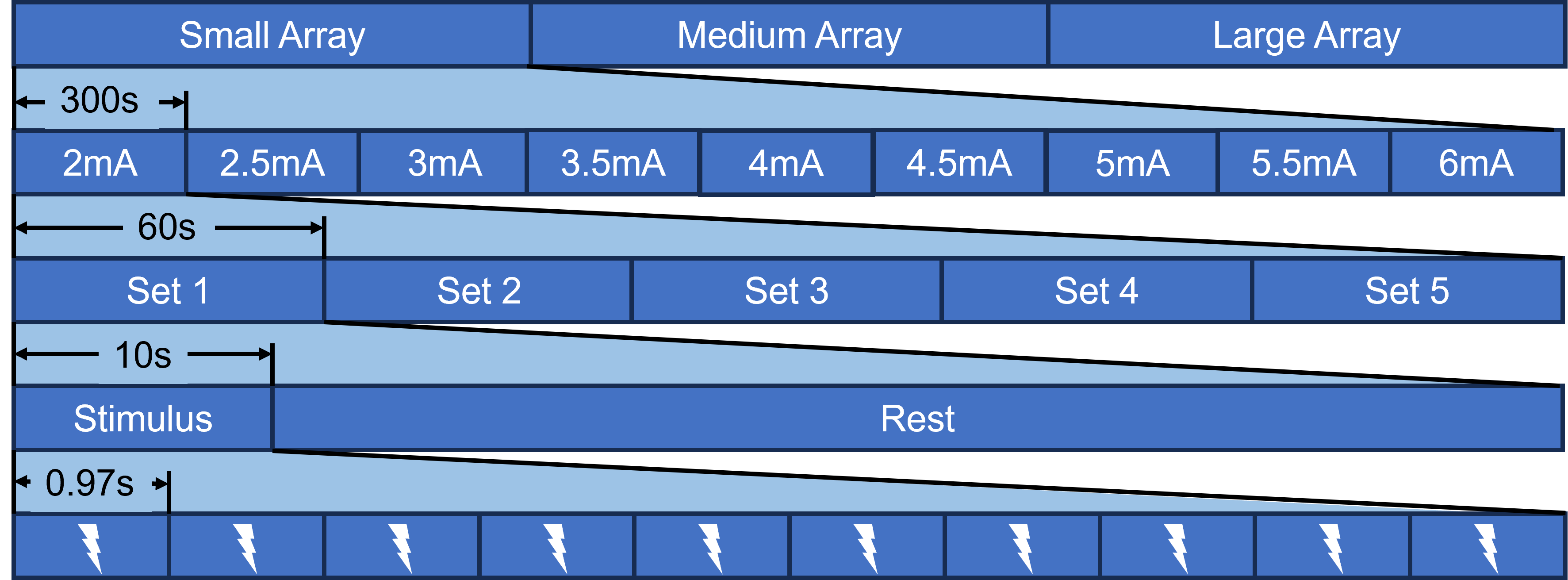}
        \caption{}
        \label{fig:CH5_Fig3_e}
    \end{subfigure} 
    
    \caption[SEP Experiment Arrays and Process]{Electrode arrays and testing procedure. (a) The electrode array used for subcutaneous stimulation to the forelimb. The (b) large (5~mm), (c) medium (3~mm), and (d) small (1~mm) custom sub-scalp arrays. The reference and ground electrodes can be seen through the substrate on the back of the arrays. These figures are the same scale. (e) A breakdown of the stimulus experiment session. Stimulus pulses were applied once a second for ten seconds, followed by 50 s of rest. This was repeated five times per stimulus amplitude for nine amplitudes about the threshold. This figure illustrates the stimulus process, for example, for a threshold amplitude of 4~mA.}
    \label{fig:CH5_Fig3}
\end{figure}

Stimulation was applied to the forelimb using a constant current stimulator (ISO Flex, A.M.P.I., Israel), controlled by a signal generator (Master 9, A.M.P.I., Israel). The stimulation threshold was determined by varying stimulation current between 0.5-10~mA in 0.5~mA intervals until muscle contraction was noticeable by eye. Stimulation was then applied to the forelimb with currents ranging between ±2~mA of this threshold, at intervals of 0.5~mA. Stimulation was applied in 1 min sets comprising ten 500 µs stimuli separated by 997 ms, followed by 50 s of rest to mitigate risk of stimulus fatigue (\Fref{fig:CH5_Fig3_e}). Stimulus current was increased after five sets resulting in 50 repetitions for each of the nine current levels. This process was repeated for the three array sizes. A trigger signal from the signal generator to the amplifier was used to annotate the exact time of stimulation.

\subsubsection{Data Preprocessing}
The data were prepared for analysis using the following process:
\begin{enumerate}
    \item Stimulus artefact removal: the stimulus creates a high amplitude artefact in the recorded data. Filtering the data without removing the stimulus artefact causes a ringing artefact that masks the stimulus response; thus, the stimulus must be removed before filtering. Due to power line noise, there is a prominent periodic signal occurring at 50~Hz. Simple linear interpolation or splining across the stimulus artefact disrupts this dominant frequency, which also leads to ringing artefact when filtering. To remove the artefact, 32 samples post-trigger were replaced with the 32 samples that occurred 96 samples prior (20~ms or one 50~Hz period).
    \item Filtering: The data was filtered with a series of 20th order, zero-phase IIR filters. These included notch filters at 50, 100, and 150~Hz to remove power line noise, and a bandpass filter between 10 and 190~Hz.
    \item Trial rejection: The data was split into epochs spanning 70~ms before and after each stimulus. Trials with a sample standard deviation greater than the 80th percentile were considered too noisy and were rejected.
    \item Re-referencing: Trials were re-referenced using common average referencing (CAR).
    \item Channel rejection: Channels were rejected by eye through observation of the average-over-trial epochs.
    \item Current threshold: The stimulus current threshold was determine as that which produced a response visually. The lowest stimulus current that resulted in a noticeable response in the average-over-trials EEG epochs for each experiment was used for analysis.
\end{enumerate}

    

\subsubsection{Signal-to-Noise Ratio (SNR)}
SNR was computed for each trial as the ratio of the variance of the data from 10-70~ms after the stimulus to the variance of the data 10-70~ms prior to the stimulus. To ensure high-quality recordings, an array should exhibit high SNR on channels near the neural activity and low SNR on distal channels. Thus, high SNR was not expected on all channels. To assess the array's capacity to record high SNR signals, the two channels with the highest SNR across experiments were compared.

\subsubsection{Cross-correlation}
Cross-correlation was used to identify common SEP spatial activity and the lag between different electrodes. Pairwise cross-correlation was calculated for average-over-trials SEPs, between channels that recorded with median SNR $>$ 0~dB and was scaled such that autocorrelations with zero lag equal to one. Maximum correlations within a $\pm$5~ms lag are reported.

\subsection{Behavioural Experiment}

The behavioural experiment to test Aim 2 closely followed previous work examining motor-related neural activity recorded with endovascular stent-electrode arrays by \citet{forsyth_evaluation_2019}.

\subsubsection{Electrode Array}
For Aim 1, cost-effective custom thin-film arrays were employed due to their configurable nature and suitability for acute experimental conditions. In contrast, Aim 2 necessitated an array optimised for chronic implantation, leading to the selection of a commercially available array designed for clinical neural recording. The electrode array (AirRay, Cortec, Germany) was a 32-channel (4$\times$8) grid array, measuring 20$\times$20~mm, with platinum electrodes ($\varnothing$~1~mm). The array was embedded in firm silicone (M4670, Barnes Products Pty Ltd, Australia) to provide anchorage points (Figure~\ref{fig:CH4_Fig1_a}). The cabling was encased within flexible tubing that was flooded with silicone (Permatex Canada Inc., Canada) to reduce the risk of interstitial fluid ingress.

\subsubsection{Implantation Procedure}
\begin{figure}
    \centering
    \begin{subfigure}[b]{0.49\textwidth}
        \includegraphics[width=\textwidth]{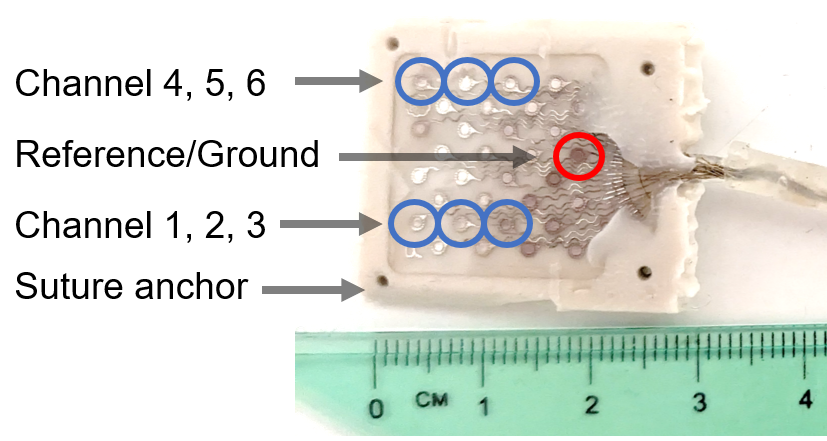}
        \caption{}
        \label{fig:CH4_Fig1_a}
    \end{subfigure}  
    \hfill
    \begin{subfigure}[b]{0.49\textwidth}
        \includegraphics[width=\textwidth]{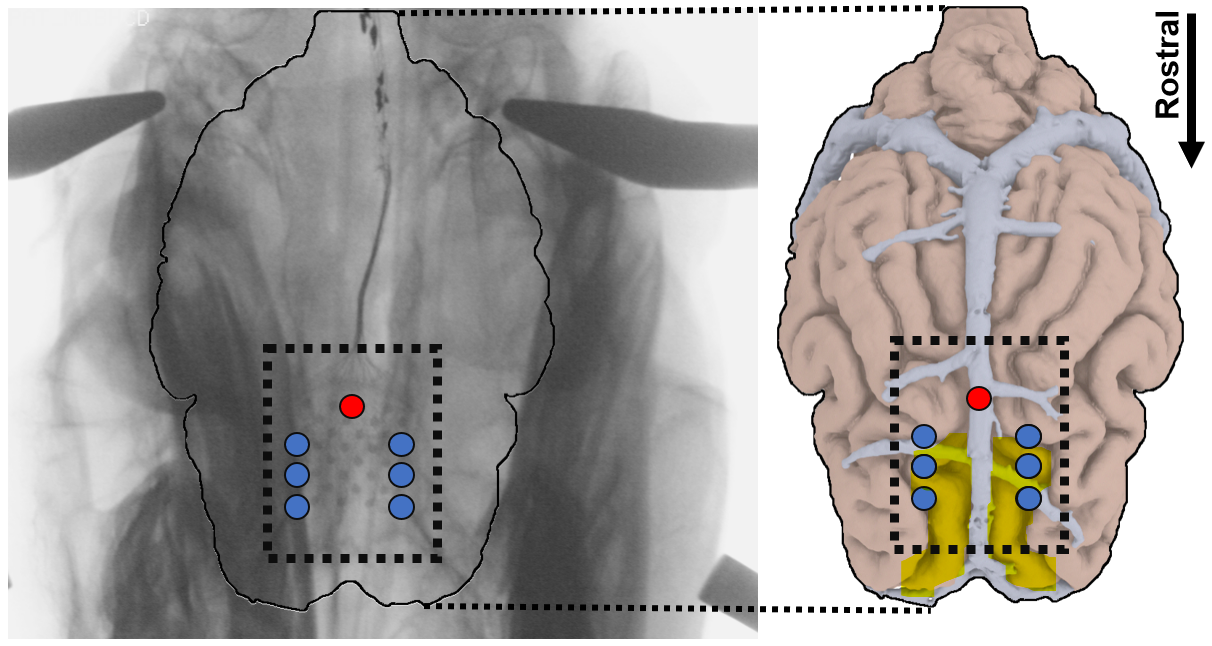}
        \caption{}
        \label{fig:CH4_Fig1_b}
    \end{subfigure} 
    \caption[Electrode Array and Location]{Electrode array and location. (a) The electrode array (AirRay, Cortec, Germany) encased in silicone (M4670, Barnes Products Pty Ltd, Australia) with anchorage points for suturing at each corner. The wires were encased within an intravenous tubing set flooded with silicone (Permatex Canada Inc., Canada). Channels are indicated by the blue circles and reference/ground by the red circle. (b) An x-ray image of the sheep head with the array (outlined by the black, dashed rectangle) implanted above the motor cortex, as indicated by the adjacent sheep brain model (source: \citet{oxley_minimally_2016}).}
\end{figure}

The sub-scalp array was implanted in four female corriedale sheep. The sheep were anaesthetised during implantation with isoflurane. Two 60 mm wide incisions were made laterally across the scalp down to the skull, one anterior and one posterior to the motor cortex. The array was sterilised in an ethanol bath prior to implantation.
The electrode array was inserted into the cavity, directly onto the skull surface. The array was sutured into place. The cabling was tunnelled from the scalp site down to the shoulder where it exited the skin through a small incision. The incision sites were then sutured closed. Behavioural experiments began one week after implantation. The device remained \textit{in vivo} for three weeks post implantation. Figure~\ref{fig:CH4_Fig1_b} is an x-ray image of an animal prior to array removal, showing the placement over the motor cortex as understood by previous ovine studies \citep{john_ovine_2017, opie_focal_2018}.

\subsubsection{Amplifier}
Unlike Aim 1, this experiment necessitated an amplifier with a lower channel count and a portable design. Portability was crucial to reduce cable length and prevent motion artifacts, ensuring the animal's unrestricted movement. As such, the amplification hardware used for the behavioural  experiment was a portible module worn on the animal, comprising an Intan RHD2032 chip (Intan Technologies, USA) that sampled six channels with 1024 Hz sampling frequency per channel. 

\subsubsection{Behavioural Experiment Setup}
\begin{figure}[ht]
\includegraphics[width=\textwidth]{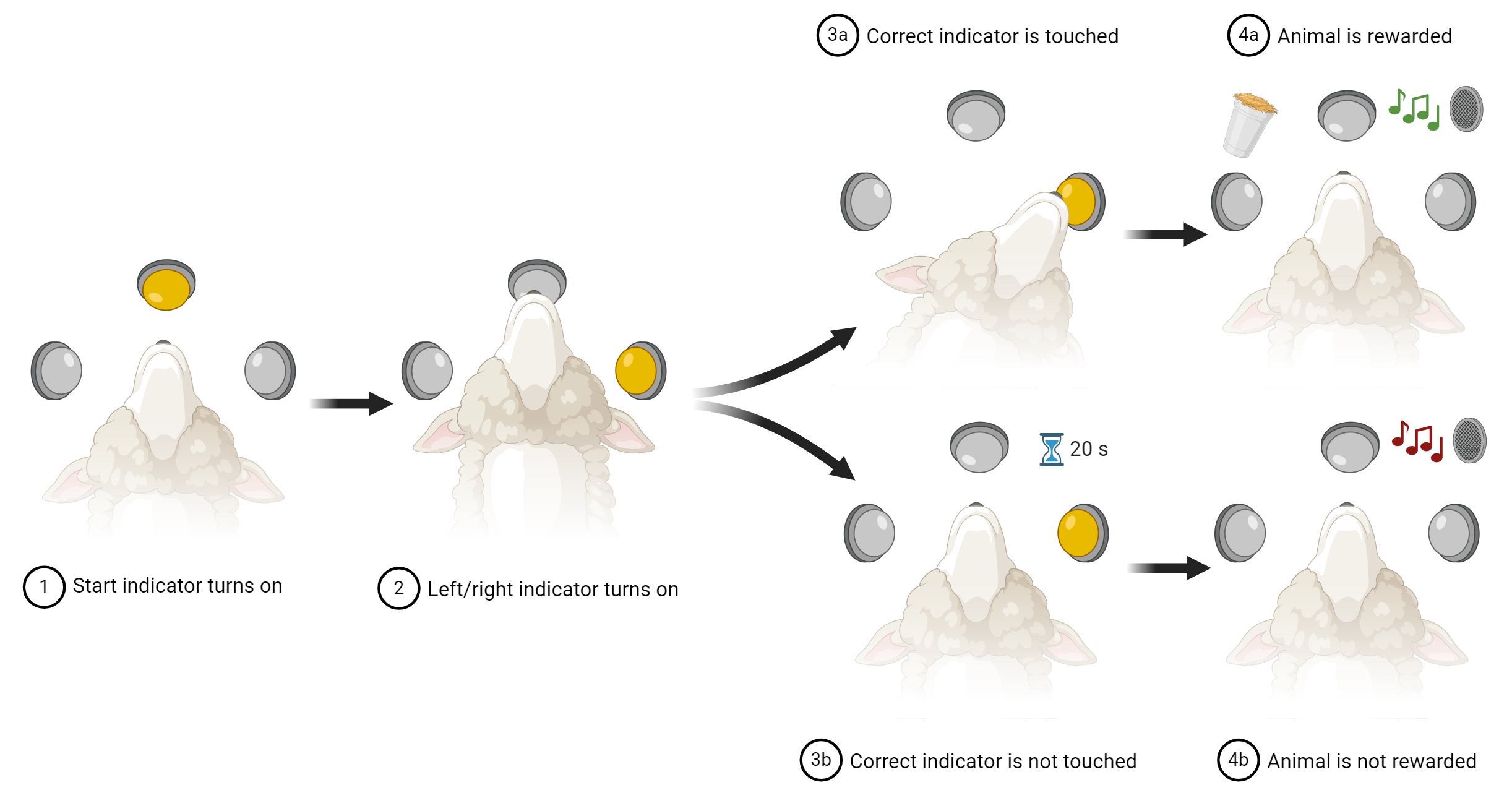}
\centering
\caption[Behavioural Experiment Process]{ The behavioural experiment progression during a single trial where the animal either successfully completes the task and is presented with a tone indicating success and a reward (3a and 4a), or fails (3b and 4b) and receives a tone indicating failure and no reward (created with BioRender.com).}
\label{fig:CH4_Fig2}
\end{figure}

The behavioural experimental setup was similar to a previous experiment performed by \citet{forsyth_evaluation_2019}, and is illustrated in Figure~\ref{fig:CH4_Fig2}. The animals were housed in individual cages. In front of each animal were three touch buttons with light emitting diodes (LEDs). The buttons were positioned at head height, within reach of the head. One button was directly in front of the animal, one was to the left, and the other was to the right. Also, in front of the animal were two audio speakers, a food trough within reach of the animal, and a camera facing the animal. 

The behavioural task was used to encourage the animal to perform left and right head movements. The animal was taught the task in a series of stages as follows:
\begin{enumerate}
    \item The equipment was set up around the animal. Food was provided in rations frequently until the animal appeared at ease with the setup and the experimenter.
    
	\item The centre button was lit up. When the animal touched the button, a single high pitched correct tone was sounded and the animal was fed a ration immediately. This continued until the animal appeared to understand the task.
 
	\item The centre button was lit up. Upon touching the centre button, one of either the left or the right buttons was lit up at random. If the animal touched the lit button within 20 seconds, a high-pitched correct tone was sounded and the animal was fed immediately. If the animal did not touch the button within the time limit, a lower pitched incorrect tone was sounded and a delay of 20 seconds prevented the animal from starting another trial. EEG data collection was started once the animal appeared to understand the task.
\end{enumerate}

The feed ration was 125~ml of a 50/50 mix of chaff/lucerne. Trials continued until either the animal lost interest or their daily allotment of food was consumed. In all cases where the animal lost interest in the task, they were fed the remainder of their daily allotment of food.

A Python script received serial triggers when the left or right buttons were pressed, as well as Bluetooth EEG data from the Intan amplifier. The camera captured the moment the animal began head movement in the direction of the button. The video combined with the timing information allowed investigation of the sub-scalp EEG during and prior to motor activity. 

\subsubsection{Analysis Procedures}
The data were bandpass filtered between 2-200 Hz and re-referenced to a common average reference. The video footage was used to determine the time that head movement began toward the target, and the data were separated into epochs spanning 200~ms prior to movement and 500 ms post movement. Spectrograms of the epochs were created using the continuous wavelet transform (MATLAB’s \texttt{cwt} function). In chronological order, every fifth epoch (20\%) was set aside as a test set. 

A filter feature selection algorithm was implemented to reduce the risk of overfitting. 
Features were selected by evaluating the mutual information between the labels of each trial and the feature set, spanning temporal, spectral, and spatial dimensions. 
The temporal feature set was restricted to 100 ms windows (no overlap) spanning the 200 ms  before and the 500 ms after movement onset. 

The frequency space comprised the bands theta [5-8 Hz], alpha [8-12 Hz], beta [12-30 Hz], gamma [30-70 Hz], and high gamma [70-200 Hz]. With the inclusion of six channels, the feature set included a total of 210 features.

Mutual information was computed by discretising the power measures using k-nearest neighbours \citep{ross_mutual_2014} with publicly available MATLAB code \citep{otoole_mutual_2020}.  The 15 features with the highest mutual information were chosen for classification. 

Classification was performed using linear discriminant analysis. We evaluated both 2-class (left vs. right) and 3-class (left vs. right vs. rest) classification cases.
Rest epochs were extracted from the data by sampling random EEG sections that did not overlap with movement epochs. Feature selection and classification was performed within a five-fold cross validation (CV) rotation. We also report the classification result of the unseen test set having trained a final model on the entire training set.
We used animal-specific feature selection and classification models to overcome inter-animal variation in feature importance. A binomial test was used to determine if classification was above chance.

\section{Results}

\subsection{SEP Stimulation Results}

For Aim 1, the experiment was performed on eight forelimbs for the 1~mm array, and seven forelimbs for the 3~mm and 5~mm arrays, across four sheep (see \tref{tab:CH5_Tab1}). Each forelimb experiment will be referred to as a session, session 1 being sheep 1, with stimulation to the left forelimb. 37\%, 22\%, and 29\% of channels for the 1~mm, 3~mm, and 5~mm arrays, respectively, were rejected by eye. 

\begin{figure}
    \centering
    \begin{subfigure}[b]{0.32\textwidth}
        \includegraphics[width=\textwidth]{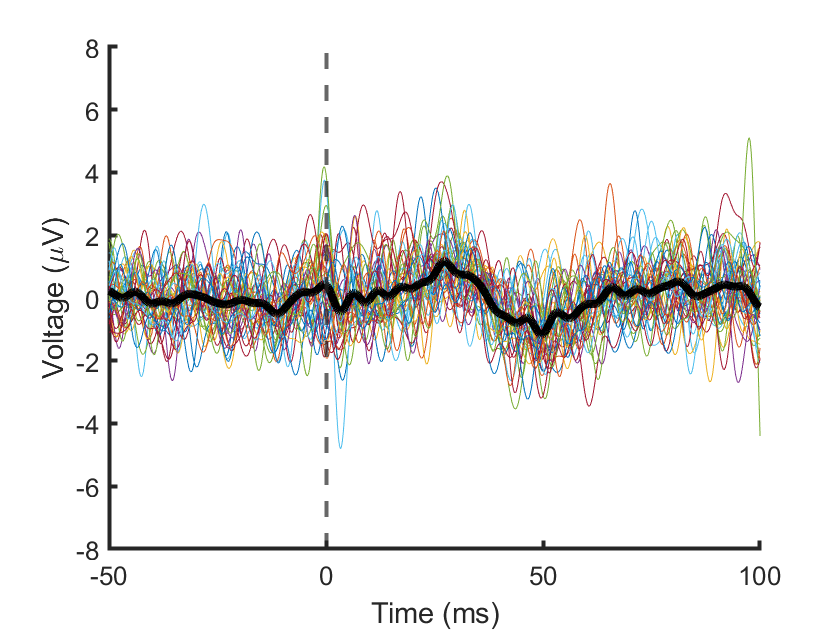}
        \caption{1mm Array}
        \label{fig:CH5_Fig5_a}
    \end{subfigure} 
    \begin{subfigure}[b]{0.32\textwidth}
        \includegraphics[width=\textwidth]{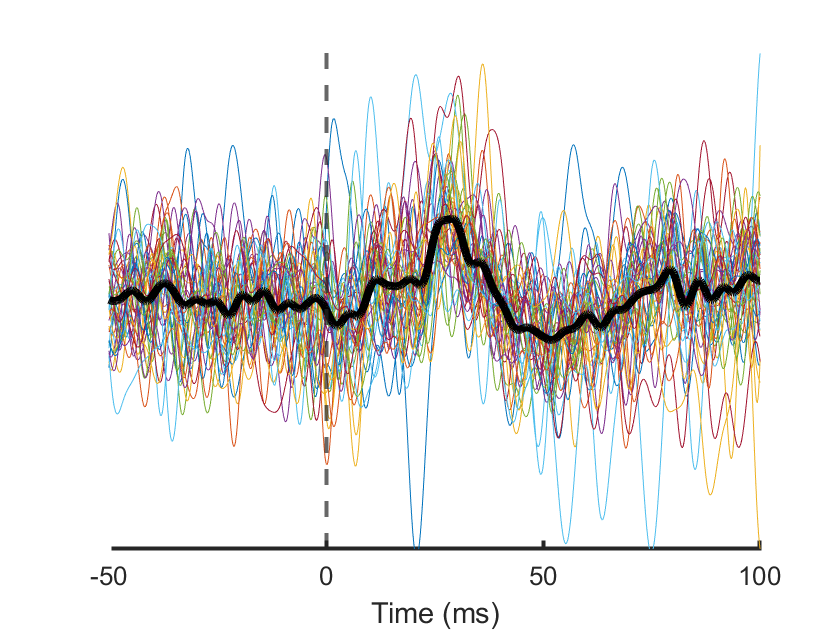}
        \caption{3mm Array}
        \label{fig:CH5_Fig5_b}
    \end{subfigure}
    \begin{subfigure}[b]{0.32\textwidth}
        \includegraphics[width=\textwidth]{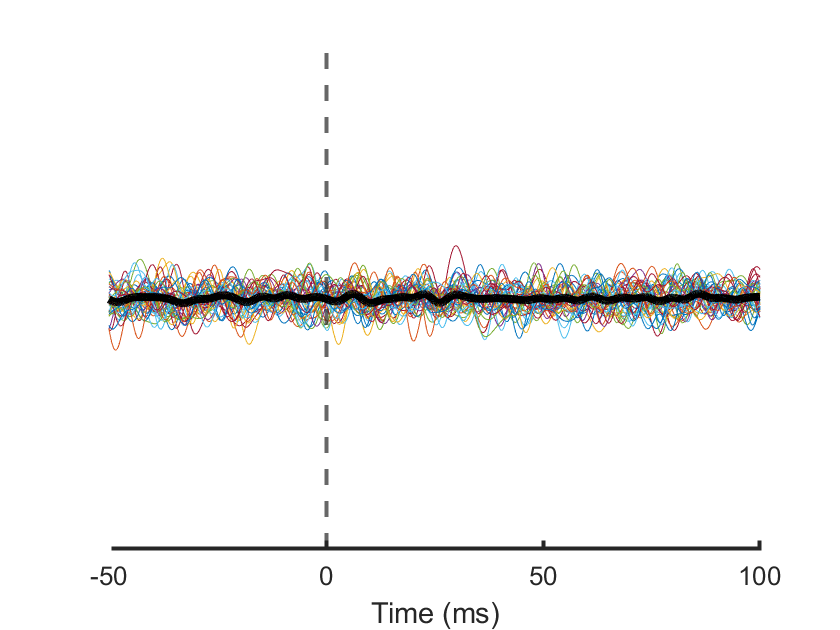}
        \caption{5mm Array}
        \label{fig:CH5_Fig5_c}
    \end{subfigure}

    \begin{subfigure}[b]{\textwidth}
        \vspace{0.5cm}
        \includegraphics[width=\textwidth]{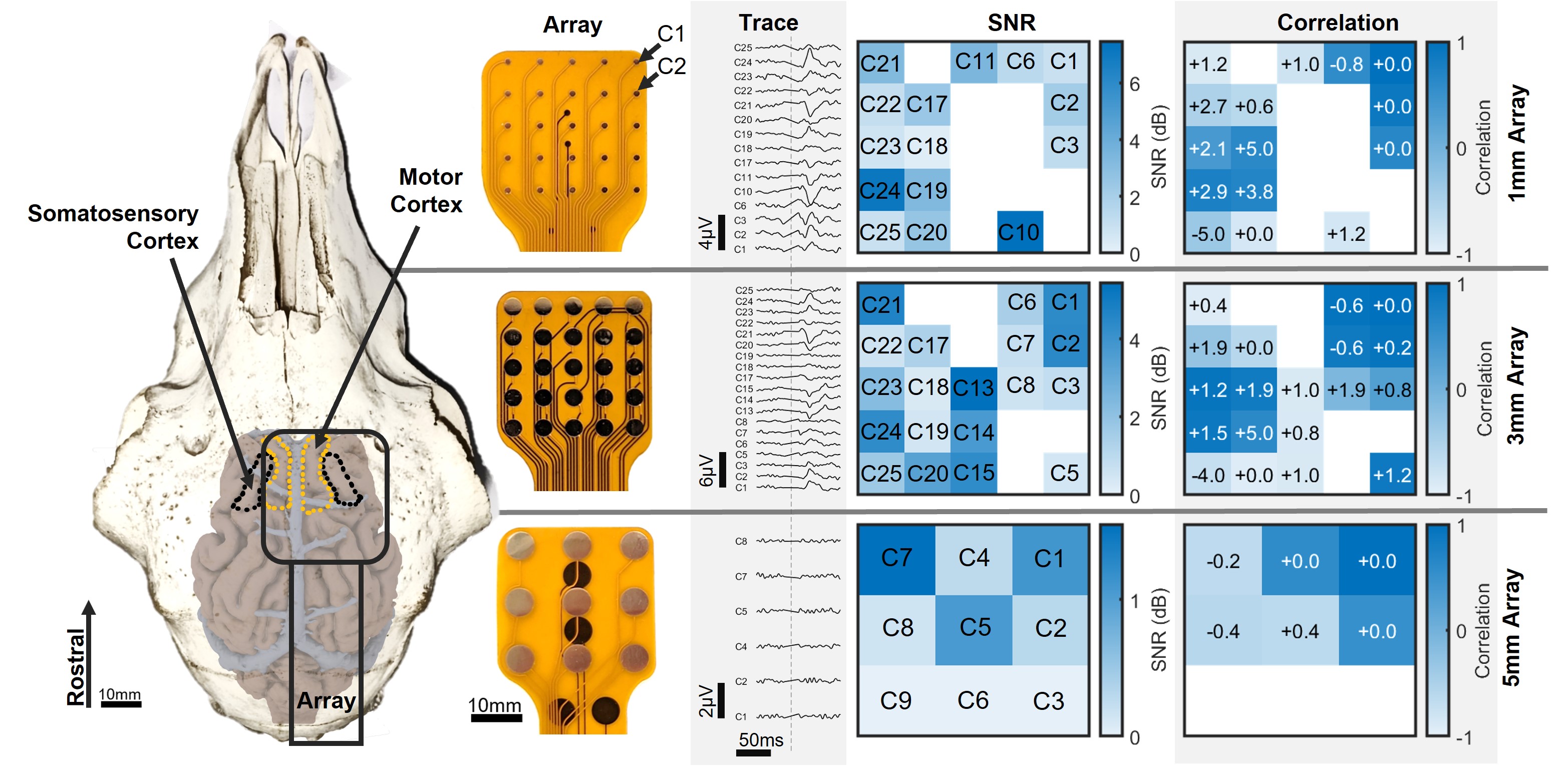}
        \caption{}
        \label{fig:CH5_Fig5_d}
    \end{subfigure} 
    \caption[SEP Responses]{SEP Responses. Exemplar single trial SEP recordings from the (a) 1~mm, (b) 3~mm, and (c) 5~mm arrays in the same animal. The average over trials is shown in bold. The dashed line indicates the time of stimulus. These figures share the same voltage scale. (d) Examples of the array placement location over the the sheep skull and approximate location of the motor and somatosensory cortices beneath, as according to \citet{john_ovine_2017}. For each array are shown exemplary average-over-trial traces, an SNR heatmap, and a correlation heatmap across channels. For the SNR heatmap, the text indicates the corresponding channel and the colour indicates the SNR amplitude. For the cross-correlation heatmap, the colour indicates the peak correlation within a $\pm$5~ms delay of channel 1, and the text indicates the correlation peak delay in milliseconds. White heatmap segments indicate either the rejection of a channel due to noise, or SNR $<$ 0~dB. This data was all collected from the same session (S5).}
\end{figure}

Figures \ref{fig:CH5_Fig5_a}, \ref{fig:CH5_Fig5_b}, and \ref{fig:CH5_Fig5_c} display exemplary single-trial SEP recordings from the 1~mm, 3~mm, and 5~mm arrays, respectively. The bold line indicates the average-over-trials SEP. In the 1~mm and 3~mm arrays, the SEP appears as a positive potential approximately 25~ms post stimulus (indicated by the dashed line), followed by a negative potential at approximately 50~ms post stimulus. SEP waveforms varied, as depicted in \Fref{fig:CH5_Fig5_d}, which displays the approximate placement of the array over the ovine sensory and motor cortices and examples of recorded averaged-over-trials SEP responses from the 1~mm, 3~mm, and 5~mm arrays. As illustrated in the traces displayed in the figure, we typically observed variations in phase and amplitude of the SEP responses across electrodes in recordings from 1~mm and 3~mm size electrodes, though not 5~mm electrodes. While previous work has noted both contralateral and ipsilateral SEP responses in sheep models \citep{john_ovine_2017}, there was no obvious laterally reflected spatial patterns observed between stimulus applied to the left forelimb versus the right forelimb. 

\Fref{fig:CH5_Fig5_d} also illustrates the spatial variation in SNR and inter-channel correlation. Within these maps, there are channels that recorded SEPs with distinct variation in SNR and correlation to adjacent channels. For example, concerning the 3~mm array, channel 19 recorded SEPs with relatively low SNR compared to adjacent channels 13, 14, 15, 20, and 24, although with similar SNR to channel 18. SNR and correlation results are discussed further in Sections \ref{sec:CH5_results_SNR} and \ref{sec:CH5_results_Corr}, respectively. For all average-over-trials traces, and SNR and correlations maps, refer to Supplementary Materials.

\begin{table}[ht]
\centering

\caption[Recording Sessions Across Sheep]{ Recording sessions (S1 to S8) across the stimulation of left (L) and right (R) forelimbs of each of the four sheep. The symbols are used to indicate sessions in subsequent figures. There were no 3~mm or 5~mm array recordings for S2 due to time constraints.}
\label{tab:CH5_Tab1}
\begin{tabular}{@{}lcccccccc@{}}
\toprule
                   & \multicolumn{8}{c}{Sheep} \\ \cmidrule(l){2-9} 
                  & \multicolumn{2}{c}{1} & \multicolumn{2}{c}{2} & \multicolumn{2}{c}{3} & \multicolumn{2}{c}{4} \\ \cmidrule(l){2-3} \cmidrule(l){4-5} \cmidrule(l){6-7} \cmidrule(l){8-9}
                  & L & R & L & R & L & R & L & R \\ \cmidrule(l){2-9}
Array Size & S1 & S2 & S3 & S4 & S5 & S6 & S7 & S8 \\ \midrule
1~mm Array         & $\bigcirc$  & + & $*$  & $\times$  & $\square$ & $\Diamond$ & \ding{73} & $\triangle$ \\
3~mm Array      & $\bigcirc$ &  & $*$  & $\times$  & $\square$ & $\Diamond$ & \ding{73} & $\triangle$ \\
5~mm Array                & $\bigcirc$ &  & $*$  & $\times$  & $\square$ & $\Diamond$ & \ding{73} & $\triangle$ \\
\bottomrule
\end{tabular}
\end{table}

\subsubsection{Signal-to-Noise Ratio (SNR)}
\label{sec:CH5_results_SNR}
\begin{figure}
    \centering
    \begin{subfigure}[b]{\textwidth}
        \includegraphics[width=\textwidth]{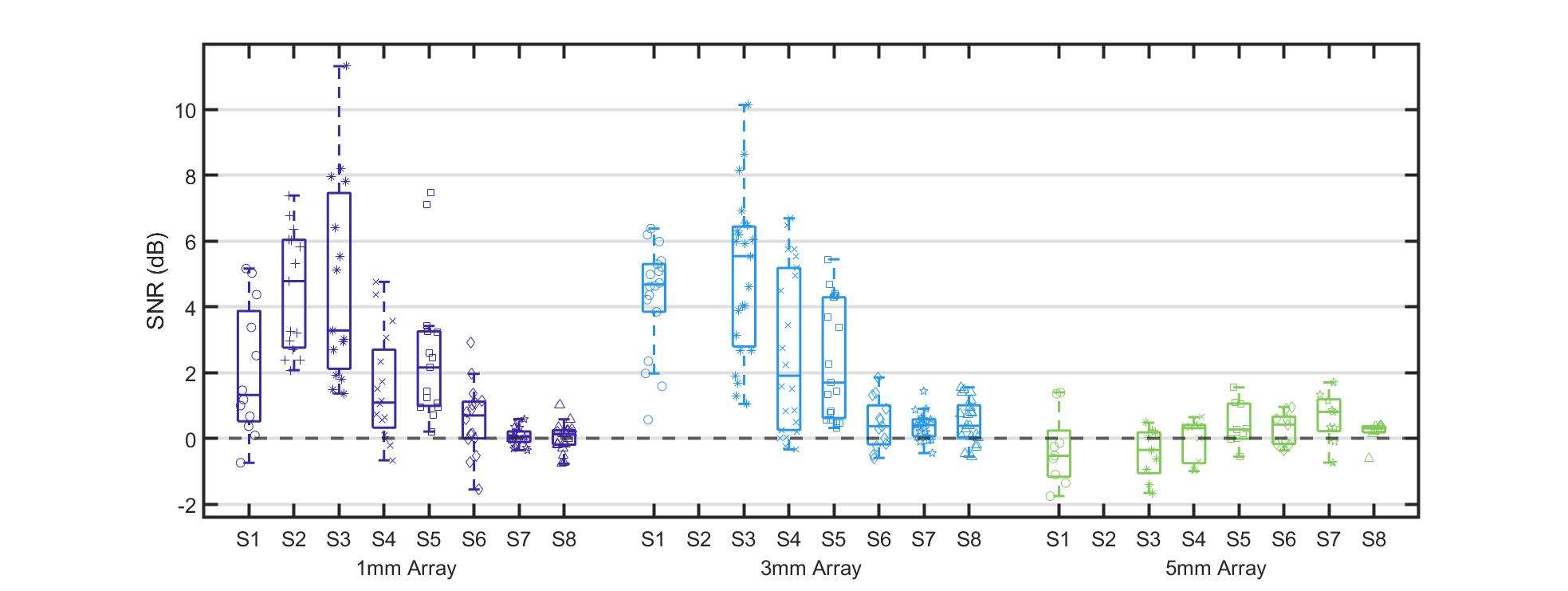}
        \caption{}
        \label{fig:CH5_Fig6_a}
    \end{subfigure} 
    
    \begin{subfigure}[b]{0.5\textwidth}
        \includegraphics[width=\textwidth]{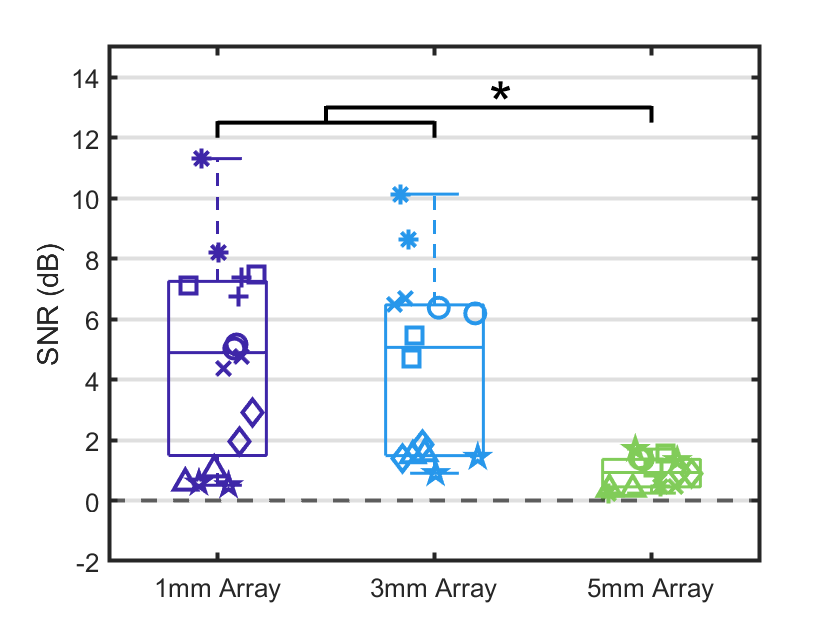}
        \caption{}
        \label{fig:CH5_Fig6_b}
    \end{subfigure}
    \caption[SEP SNR Across Arrays]{ (a) The median SEP SNR for each channel across sessions and arrays. Each marker is the median SEP SNR across trials for a single channel. S1 to S8 are the different sessions. (b) The two channels with highest SNR grouped across sessions for each array. Each marker type indicates the recording session. The significance marker ($*$) indicates a p-value less than 0.01.}
\end{figure}

\Fref{fig:CH5_Fig6_a} shows the spread of SNR across channels for each array and session. The median SNR across channels was significantly greater than 0~dB in 5 of the 8 sessions (62.5\%) with the 1~mm array, 6 of the 7 sessions (85.7\%) with the 3~mm array, and 0 of the 7 sessions (0\%) with the 5~mm array ($\alpha=0.01$, Student's t-test). P-values for each session are provided in \tref{tab:CH5_Tab2}. SNR varied more between channels in sessions recorded with the 1~mm array ($\sigma^2$=3.1$\pm$3.0~$\mu$V\textsuperscript{2}, median$\pm$std across sessions) and 3~mm array ($\sigma^2$=2.7$\pm$2.6~$\mu$V\textsuperscript{2}) recordings compared to the 5~mm array ($\sigma^2$=0.4$\pm$0.35~$\mu$V\textsuperscript{2}). 

\Fref{fig:CH5_Fig6_b} is a comparison of the two channels with the highest SNR for each session, across array types. One-way ANOVA revealed a significant difference between the groups, F(2, 41) = 9.48, p=0.0004. Post-hoc analysis (Tukey's HSD) indicated that the top-two SEP SNRs recorded from 1~mm and 3~mm electrodes were significantly greater than those recorded using 5~mm electrodes (p=0.0006 and p=0.0017, respectively, $\alpha=0.01$). There was no significant difference between SNR from 1~mm and 3~mm electrodes (p=0.9659). All groups in \Fref{fig:CH5_Fig6_b} recorded SEP SNR significantly greater than 0~db  (p$<0.0001$, $\alpha=0.01$,  Student's t-test).

\begin{table}[ht]
\centering

\caption[SNR Student's t-test Results]{ The Student's t-test p-values for each array and session to test if SNR is significantly greater than zero across all channels. We also include the percentage of sessions that recorded with SNR significantly greater than zero.}
\label{tab:CH5_Tab2}
\begin{tabular}{@{}crrrrrr@{}}
\toprule
   & \multicolumn{6}{c}{Array} \\ \cmidrule(l){2-7} 
  & \multicolumn{2}{c}{1~mm} & \multicolumn{2}{c}{3~mm} & \multicolumn{2}{c}{5~mm} \\ \cmidrule(l){2-3} \cmidrule(l){4-5} \cmidrule(l){6-7}
Session  & \multicolumn{1}{c}{p} & $<\alpha$ & \multicolumn{1}{c}{p} & $<\alpha$ & \multicolumn{1}{c}{p} & $<\alpha$ \\ \midrule
S1 & 0.0048 & true & $<0.0001$ & true & 0.3958 & false  \\
S2 & $<0.0001$ & true & &  &  &  \\
S3 & $<0.0001$ & true & $<0.0001$ & true & 0.1136 & false \\
S4 & 0.0020 & true & 0.001 & true & 0.8236 & false \\
S5 & 0.0005 & true & $<0.0001$ & true & 0.0910 & false \\
S6 & 0.0481 & false & 0.0536 & false & 0.1182 & false  \\
S7 & 0.4011 & false & 0.0004 & true & 0.0363 & false \\
S8 & 0.7796 & false & 0.0017 & true & 0.2304 & false \\ \midrule
\% p $<\alpha$ & \multicolumn{2}{c}{62.5\%} & \multicolumn{2}{c}{85.7\%} & \multicolumn{2}{c}{0.0\%} \\ 
\bottomrule
\end{tabular}
\end{table}

\subsubsection{Cross-Correlation}
\label{sec:CH5_results_Corr}
\begin{figure}[htb]
    \centering
    \begin{subfigure}[b]{\textwidth}
        \includegraphics[width=\textwidth]{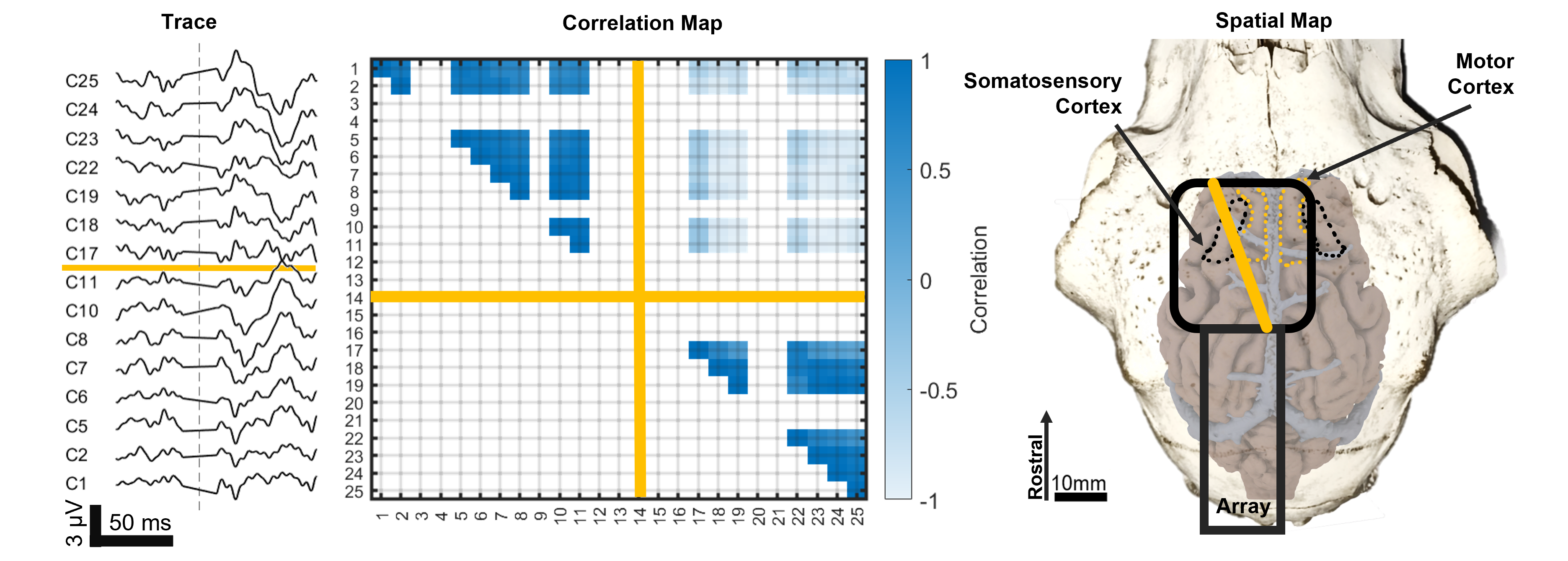}
        \caption{S2, 1~mm Array}
        \label{fig:CH5_Fig7_a}
    \end{subfigure} 
    
    \begin{subfigure}[b]{0.302\textwidth}
        \includegraphics[width=\textwidth]{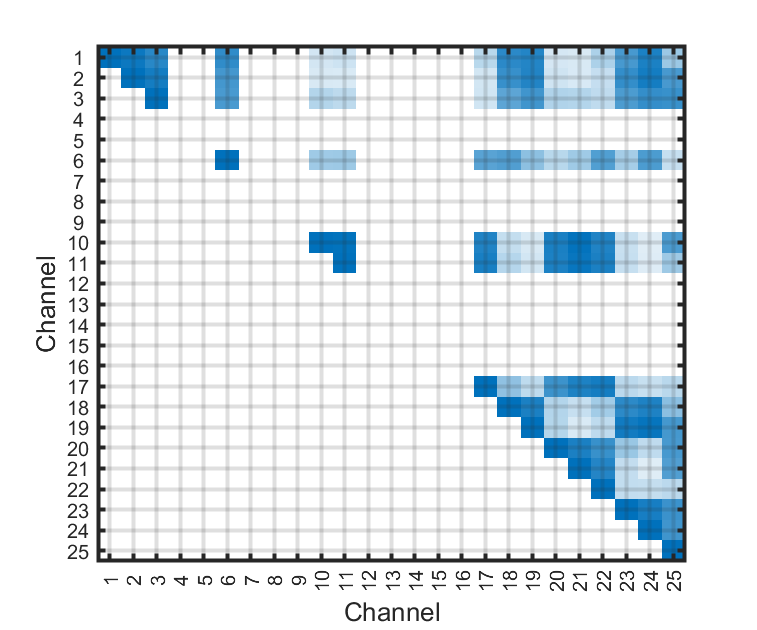}
        \caption{S5, 1~mm Array}
        \label{fig:CH5_Fig7_b}
    \end{subfigure} 
    \begin{subfigure}[b]{0.302\textwidth}
        \includegraphics[width=\textwidth]{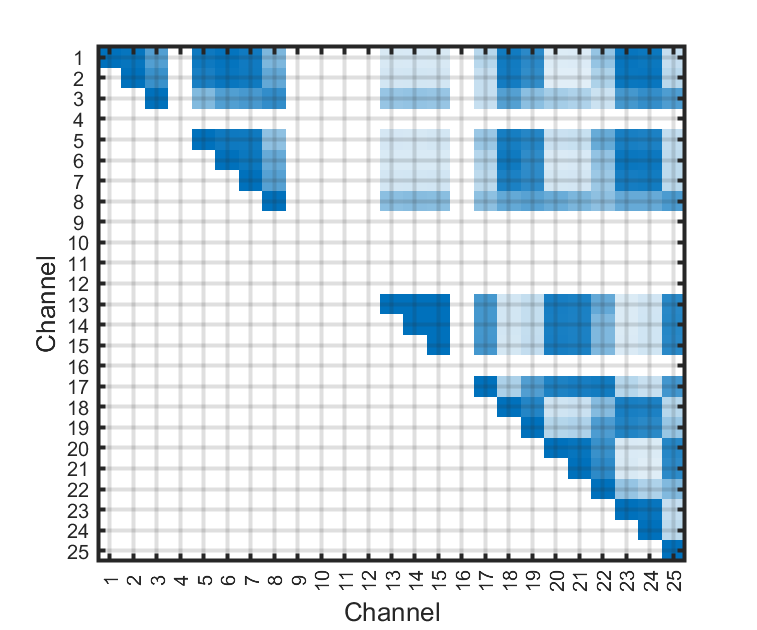}
        \caption{S5, 3~mm Array}
        \label{fig:CH5_Fig7_c}
    \end{subfigure}
    \begin{subfigure}[b]{0.336\textwidth}
        \includegraphics[width=\textwidth]{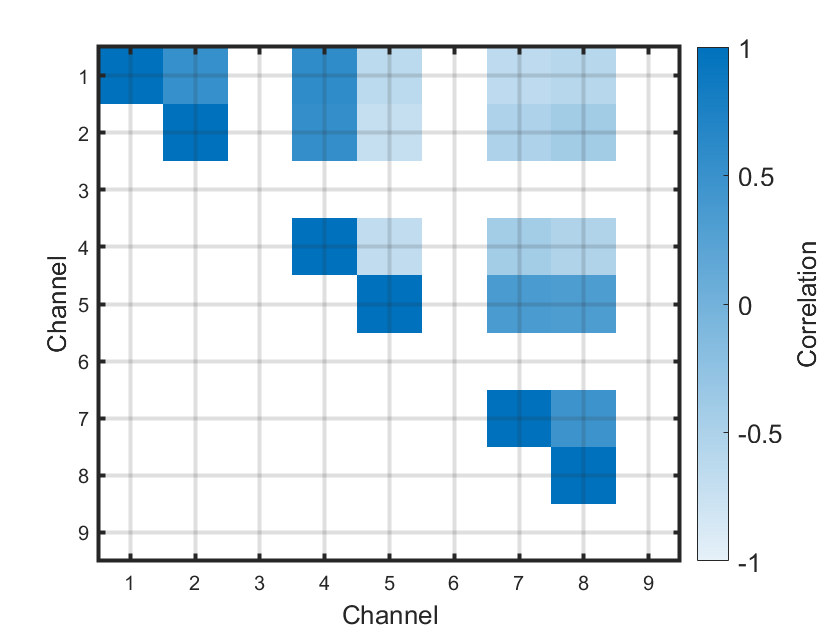}
        \caption{S5, 5~mm Array}
        \label{fig:CH5_Fig7_d}
    \end{subfigure}

    \caption[Pairwise Channel Correlation Maps and Histograms]{Pairwise channel correlation maps and histograms. (a) Phase variation across channels from one session (S2, 1~mm array). The traces on the left are the average-over-trial SEPs. The dashed line indicates the time of stimulus. The orange line illustrates the change in SEP phase between channels 1 to 11 and 17 to 25. This change in phase is consistent with computed inter-channel correlation values as per the adjacent heatmap, which depicts the pairwise channel correlations as a top-diagonal matrix. The map confirms that channels 1 to 11 and 17 to 25 form two groups comprising highly correlated channels, and that the groups are negatively correlated with each other. Similarly to the trace figure, the orange line between channels 11 and 17 discriminates between the two groups. On the right, the orange line is indicated spatial on the array, which is overlayed in the approximate location of recording. SEP (b), (c), (d) Examples of other correlation maps for each array from a single session (S5). White segments indicate either rejected channels or channels with SNR less than 0~dB.} 
\end{figure}

Cross-correlation varied across channels in each session and array. There were groups of adjacent channels with high positive correlation and groups of adjacent channels with high negative correlation. \Fref{fig:CH5_Fig7_a} depicts the change in phase across channels from session 2, as recorded with the 1~mm array. This example has been highlighted as the phase change depicted two clearly separable groups, as illustrated by the orange line. In this example, channels 1 to 11 are correlated, forming a group on the correlation map. Although there is variation between these channels, generally they depict an SEP with a negative potential followed by a positive potential. Channels 17 to 25 are also positively correlated, and are negatively correlated to channels 1 to 11. This relationship can be observed in the traces, where the SEP of these channels is instead a positive potential followed by a negative potential. Typically, as depicted in Figures \ref{fig:CH5_Fig7_b} to \ref{fig:CH5_Fig7_d} and further in Supplementary Materials, correlations were clustered in smaller, disconnected groups. 

\clearpage

\subsection{Behavioural Results}

\begin{figure}
    \begin{subfigure}[b]{0.38\textwidth}
        \includegraphics[width=\textwidth]{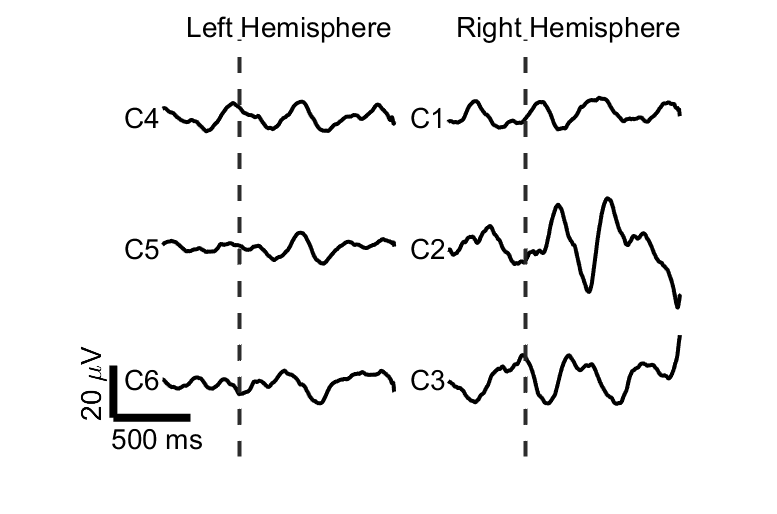}
        \caption{Left Move Traces}
        \label{fig:CH4_Fig4_a}
    \end{subfigure}  
    \begin{subfigure}[b]{0.38\textwidth}
        \includegraphics[width=\textwidth]{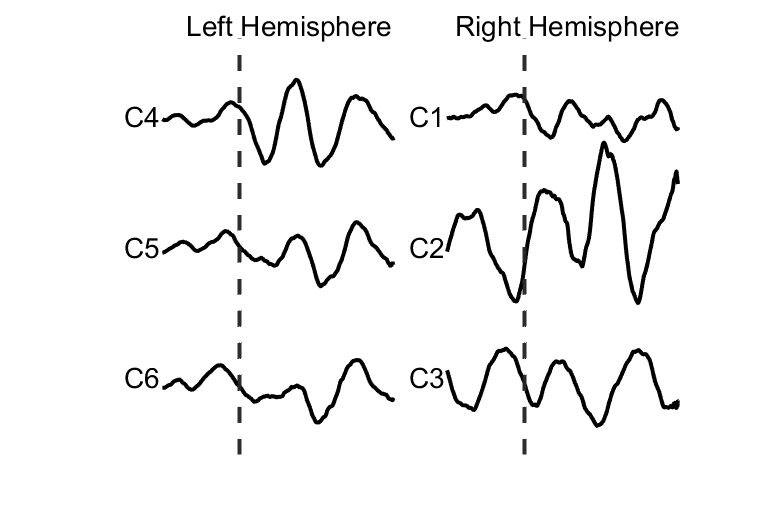}
        \caption{Right Move Traces}
        \label{fig:CH4_Fig4_b}
    \end{subfigure} 

    \sbox0{\includegraphics[width=0.39\textwidth]{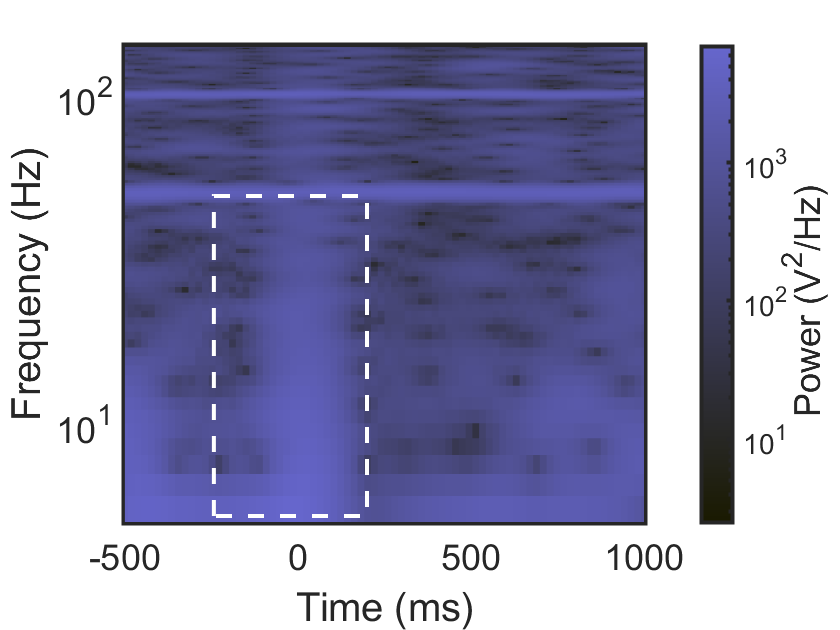}}
    \sbox1{\includegraphics[width=0.185\textwidth]{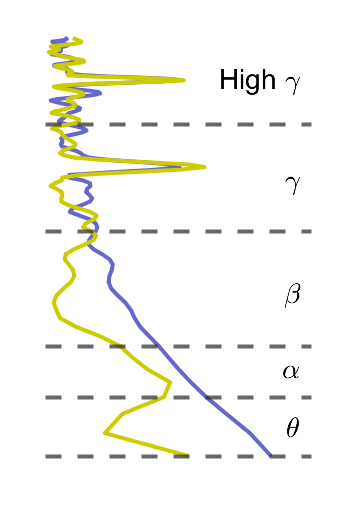}}
    
    \begin{subfigure}{\wd0}
        \includegraphics[width=\textwidth]{CH4_Fig4_c.png}
        \caption{Left Move Spectrum}
        \label{fig:CH4_Fig4_c}
    \end{subfigure}  
    \begin{subfigure}{\wd0}
        \includegraphics[width=\textwidth]{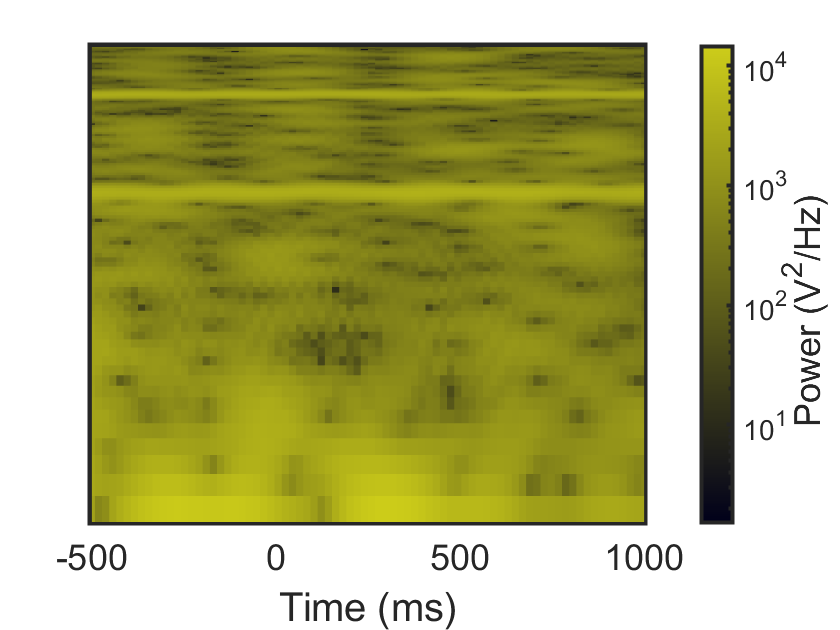}
        \caption{Right Move Spectrum}
        \label{fig:CH4_Fig4_d}
    \end{subfigure}
    \begin{subfigure}{\wd1}
        \usebox1
        \vspace{\dimexpr \ht1-\ht0}
        \caption{}
        \label{fig:CH4_Fig4_e}
    \end{subfigure} 
    \caption[Traces and Spectrograms for Left and Right Movement]{ The averaged-over-trials traces for (a) left and (b) right movements in sheep 1. These traces were lowpass filtered with a cutoff of 30 Hz to help visualise the event related potentials. The dashed lines indicate the time of movement onset. The averaged-over-trials spectrograms of the (c) left and (d) right movement of sheep 1, as recorded by channel 5 (C5). The time scale is relative to movement onset. (e) The mean log-power spectra during the 0-100 ms window post stimulus. The line colour indicates the corresponding spectrum (blue - left movement, yellow - right movement). EEG frequency bands are indicated by the dashed lines.}
    \label{fig:CH4_Fig4}
\end{figure}

For Aim 2, arrays were implanted in four sheep. All four animals successfully learned the task. At least 240 trials (120 left, 120 right) were recorded from each sheep across multiple sessions. Recording examples from sheep 1 are shown in Figure~\ref{fig:CH4_Fig4}. 

Figures~\ref{fig:CH4_Fig4_a} and \ref{fig:CH4_Fig4_b} are the averaged-over-epochs EEG traces from sheep 1, depicting motor event related potentials. The potentials showed variations in amplitude and phase between left and right movements. 

The frequency space revealed additional distinctions between left and right movements. Figures~\ref{fig:CH4_Fig4_c} and \ref{fig:CH4_Fig4_d} are the mean spectrograms over all trials from sheep 1, channel 5, for left and right movements, respectively. As indicated by the dashed rectangle, there was a clear increase in power (synchronisation) across the theta, alpha, and beta frequency bands during left movement onset (Figure~\ref{fig:CH4_Fig4_c}) that was not recorded during right movement onset (Figure~\ref{fig:CH4_Fig4_d}). A comparison of the mean power spanning 0 to 100~ms post movement onset between left (blue) and right (yellow) movements is illustrated in Figure~\ref{fig:CH4_Fig4_e}. These findings demonstrate that useful features for movement classification were spread across different time windows, frequency bands, and channels. 

\subsection{Classification}
\label{sec:CH4_classification}
\Tref{tab:CH4_Tab1} shows the classification results for both the 5-fold CV and test set models
, as well as the number of trials tested on ($n$), the resulting p-value, and the sensitivity and specificity for each class. For the two-class case of left vs. right movement classification, above chance performance (p$<$0.05) was achieved in two of the four animals, both in five-fold CV and the test set. Despite performing above chance in the test set (p=0.032), Sheep 2 classification accuracy was not considered above chance given the poor CV performance (p=0.436). When temporal features were limited to those prior to movement, performance was not above chance for any animals. 

Considering the three-class case of left vs. right vs. rest classification, above chance performance was achieved in three of the four sheep in both CV and test sets. 
Data collected from sheep 4 failed to perform above chance level in any case.

\pagebreak
\begin{table}[htb]
\centering
\caption[Classification Accuracy Results]{ Classification accuracy results of five-fold CV (mean$\pm$std) and test set prediction for the two-class (left vs. right) and three-class (left vs. right vs. rest) cases. For each case, p-values and test set sizes ($n$) are listed. Accuracy results in bold are significantly above chance (p$<$0.05). Classes were equally frequent in both training and test sets; hence chance level is 50\% for the two-class case, and 33.$\dot{3}$\% for the three-class case.}
\label{tab:CH4_Tab1}
\footnotesize
\begin{tabular}{ccccclcccccc}
\toprule
\multicolumn{12}{c}{\textbf{Left vs. Right}}  \\ \midrule
                   & \multicolumn{3}{c}{5-Fold Cross Validation}   & \multicolumn{8}{c}{Test Set} \\ \cmidrule{2-4} \cmidrule(l){5-12}
                   &                      & \multicolumn{2}{c}{Accuracy}                       &                     &       & \multicolumn{2}{c}{Sensitivity} & \multicolumn{2}{c}{Specificity} & \multicolumn{2}{c}{Accuracy} \\ \cmidrule(l){3-4} \cmidrule(l){7-8} \cmidrule(l){9-10} \cmidrule(l){11-12}
Sheep              & n                    & \%                     & p-value            & n                   & Class & \%          & p-value           & \%          & p-value           & \%                  & p-value  \\ \midrule

\multirow{2}{*}{1} & \multirow{2}{*}{42}  & \multirow{2}{*}{\textbf{65$\pm$15}} & \multirow{2}{*}{\phantom{$<$}0.021}    &  \multirow{2}{*}{54} & Left  & 89          & $<$0.001          & 44          & \phantom{$<$}0.752             & \multirow{2}{*}{\textbf{66}} & \multirow{2}{*}{\phantom{$<$}0.005}    \\ 
                   &                           &                        &                        &                     & Right & 44          & \phantom{$<$}0.752             & 89          & $<$0.001          &                     &\\ \arrayrulecolor{black!10}\midrule
                   
\multirow{2}{*}{2} & \multirow{2}{*}{38}  &  \multirow{2}{*}{51$\pm$8\phantom{0}}  & \multirow{2}{*}{\phantom{$<$}0.436} & \multirow{2}{*}{50} & Left  & 80          & $<$0.001          & 44          & \phantom{$<$}0.760              & \multirow{2}{*}{\textbf{62}} & \multirow{2}{*}{\phantom{$<$}0.032} \\
                   &                           &                        &                        &                     & Right & 44          & \phantom{$<$}0.760              & 80          & $<$0.001          &                     &      \\ \arrayrulecolor{black!10}\midrule

\multirow{2}{*}{3} & \multirow{2}{*}{48}  & \multirow{2}{*}{\textbf{67$\pm$9}\phantom{0}}  & \multirow{2}{*}{\phantom{$<$}0.007}    & \multirow{2}{*}{62} & Left  & 68          & \phantom{$<$}0.002             & 65          & \phantom{$<$}0.008             & \multirow{2}{*}{\textbf{66}} & \multirow{2}{*}{\phantom{$<$}0.004}    \\
                   &                           &                        &                        &                     & Right & 65          & \phantom{$<$}0.008             & 68          & \phantom{$<$}0.002             &                     &                           \\ \arrayrulecolor{black!10}\midrule
\multirow{2}{*}{4} & \multirow{2}{*}{46}  & \multirow{2}{*}{56$\pm$2\phantom{0}}  & \multirow{2}{*}{\phantom{$<$}0.231}    & \multirow{2}{*}{60} & Left  & 63          & \phantom{$<$}0.014             & 30          & \phantom{$<$}0.999             & \multirow{2}{*}{47} & \multirow{2}{*}{\phantom{$<$}0.651}     \\
                  &                           &                        &                        &                     & Right & 30          & \phantom{$<$}0.999             & 63          & \phantom{$<$}0.014             &                     &      \\ \arrayrulecolor{black}\toprule
                   
\multicolumn{12}{c}{\textbf{Left vs. Right vs. Rest}}   \\ \midrule
                   & \multicolumn{3}{c}{5-Fold Cross Validation}   & \multicolumn{8}{c}{Test Set}  \\ \cmidrule{2-4} \cmidrule(l){5-12}
                   &                      & \multicolumn{2}{c}{Accuracy}     &      &       & \multicolumn{2}{c}{Sensitivity} & \multicolumn{2}{c}{Specificity} & \multicolumn{2}{c}{Accuracy}  \\ \cmidrule(l){3-4} \cmidrule(l){7-8} \cmidrule(l){9-10} \cmidrule(l){11-12} 
                   
Sheep              & n                    & \%                     & p-value            & n                   & Class & \%          & p-value           & \%          & p-value           & \%                  & p-value    \\ \midrule

\multirow{3}{*}{1} & \multirow{3}{*}{63}  & \multirow{3}{*}{\textbf{57$\pm$10}} & \multirow{3}{*}{$<$0.001} & \multirow{3}{*}{81} & Left  & 85          & $<$0.001          & 37          & \phantom{$<$}0.204             & \multirow{3}{*}{\textbf{53}} & \multirow{3}{*}{$<$0.001} \\
                   &                           &                        &                        &                     & Right & 33          & \phantom{$<$}0.448             & 63          & $<$0.001          &                     &  \\
                   &                           &                        &                        &                     & Rest  & 41          & \phantom{$<$}0.065             & 59          & $<$0.001          &                     & \\ \arrayrulecolor{black!10}\midrule
                   
\multirow{3}{*}{2} & \multirow{3}{*}{57}  & \multirow{3}{*}{\textbf{46$\pm$7}\phantom{0}}  & \multirow{3}{*}{\phantom{$<$}0.019}    &  \multirow{3}{*}{75} & Left  & 52          & $<$0.001          & 56          & $<$0.001          & \multirow{3}{*}{\textbf{55}} & \multirow{3}{*}{$<$0.001}  \\

                   &                           &                        &                        &                     & Right & 44          & \phantom{$<$}0.020             & 60          & $<$0.001          &                     &                            \\
                   &                           &                        &                        &                     & Rest  & 68          & $<$0.001          & 48          & \phantom{$<$}0.003             &                     &                           \\ \arrayrulecolor{black!10}\midrule
\multirow{3}{*}{3} & \multirow{3}{*}{72}  & \multirow{3}{*}{\textbf{52$\pm$13}} & \multirow{3}{*}{$<$0.001} & \multirow{3}{*}{93} & Left  & 61          & $<$0.001          & 63          & $<$0.001          & \multirow{3}{*}{\textbf{62}} & \multirow{3}{*}{$<$0.001} \\
                   &                           &                        &                        &                     & Right & 58          & $<$0.001          & 65          & $<$0.001          &                     &                           \\
                   &                           &                        &                        &                     & Rest  & 68          & $<$0.001          & 60          & $<$0.001          &                     &                           \\ \arrayrulecolor{black!10}\midrule
\multirow{3}{*}{4} & \multirow{3}{*}{69}  & \multirow{3}{*}{36$\pm$6\phantom{0}}  & \multirow{3}{*}{\phantom{$<$}0.347}  & \multirow{3}{*}{90} & Left  & 67          & $<$0.001          & 17          & \phantom{$<$}1.000                 & \multirow{3}{*}{33} & \multirow{3}{*}{\phantom{$<$}0.540} \\
                  &                           &                        &                        &                     & Right & 10          & \phantom{$<$}1.000                 & 45          & \phantom{$<$}0.011             &                     &   \\
                   &                           &                        &                        &                     & Rest  & 23          & \phantom{$<$}0.974             & 38          & \phantom{$<$}0.157             &                     & \\ \arrayrulecolor{black}\bottomrule  
\end{tabular}
\end{table} 

\begin{figure}[ht]
\includegraphics[width=\textwidth]{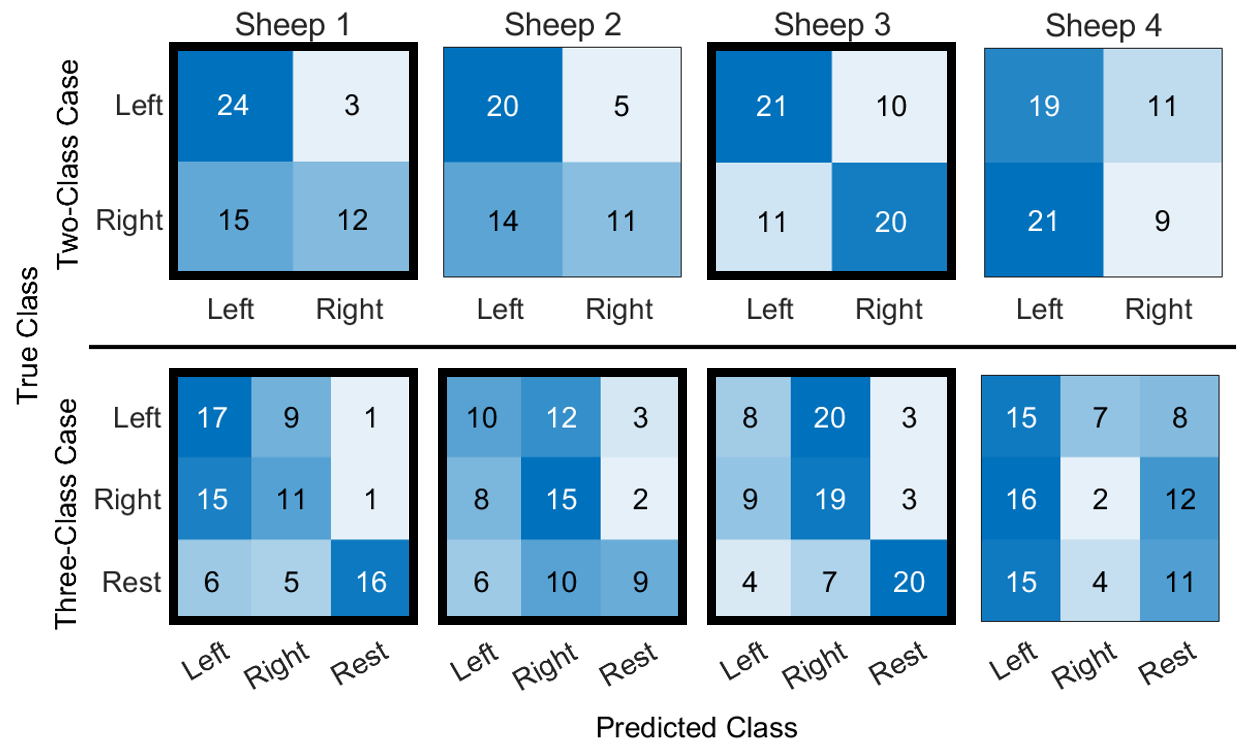}
\centering
\caption[Three-class Classification Confusion Matrices]{ Confusion matrices for each classification case. The matrices with a bold outline indicate cases that resulted in above chance classification accuracy.}
\label{fig:CH4_Fig7}
\end{figure}

Figure~\ref{fig:CH4_Fig7} shows the confusion matrices for each classification case test set, with cases resulting in above chance classification highlighted. Sheep 1 and 3 performed above chance in the two-class case, with sensitivity and specificity of 89\% and 44\%, and 68\% and 65\% respectively (left=positive, right=negative). For the three-class cases, the rest state had high mean class sensitivity for the motor execution case (74\%) compared to the left and right classes (47\% and 61\%, respectively). 

\clearpage

\section{Discussion}

\subsection{Aim 1: Sub-scalp EEG Spatial Resolution}
Understanding the spatial resolution of EEG recorded from the sub-scalp space can provide supporting evidence for ventures seeking to develop sBCI systems for use with patients. The first aim of this study was to investigate the spatial resolution of sub-scalp EEG. We successfully recorded and analysed SEP responses from the sub-scalp space in animals models with arrays of varying electrode size. The results provide insight through analysis of spatial variations in SNR and correlation across channels. Our findings indicate that a 5~mm inter-channel distance reveals distinct spatial information, and that arrays with 1~mm and 3~mm electrode diameter record SEPs with higher SNR and inter-channel variance than arrays with 5~mm electrode diameter. Here, we will discuss the variations in SNR and correlation and relate them to spatial resolution, compare with alternative BCI signal acquisition methods, and note limitations of the work.

\subsubsection{Signal-to-Noise Ratio}
The 1~mm and 3~mm diameter electrodes recorded SEPs with significantly higher SNR (4.9$\pm$3.2~dB and 5.1$\pm$3~dB (median$\pm$std), respectively) than the 5~mm electrodes (0.9$\pm$0.5~dB).  Larger electrodes record activity from broader areas. The 5~mm electrodes may have captured signals from such a large cortical area that their sensitivity regions overlapped, resulting in low SNR after common average referencing. While increasing the inter-channel distance may record less correlated signals, this would come at the cost of reduced spatial resolution.

The SNR values are in a similar range to previous work by \citet{john_signal_2018}, who also investigated SEP response in sheep, reporting single trial SNR from ECoG (epidural: $\sim$5~dB, subdural: $\sim$8~dB) and endovascular ($\sim$5~dB) arrays. Visual evoked potentials in sheep models have been recorded from endovascular arrays with similar single trial SNRs of 4.35 \citep{gerboni_visual_2018} and 4.77~dB (reported as 3.05~V\textsuperscript{2}/V\textsuperscript{2}) \citep{mahoney_comparison_2023}. However, direct comparison with these results is limited due to variations in the method of stimulation, hardware, data processing, and SNR calculation. Additionally, \citet{john_signal_2018} allowed the electrodes to settle in the body for 3-4 weeks prior to recording, which would likely improve electrode contact and recording quality. Despite these differences in methods, recording SEP SNR with comparably high SNR to more invasive ECoG and endovascular arrays suggests sub-scalp arrays may record with similarly high signal quality.

SNR also showed higher variance between channels with the 1~mm and 5~mm arrays (3.1$\pm$3.0~$\mu$V\textsuperscript{2} and 2.7$\pm$2.6~$\mu$V\textsuperscript{2}, respectively) than the 5~mm arrays (0.4$\pm$0.3~$\mu$V\textsuperscript{2}). High variance may indicate that spatial resolution is high enough to isolate neural regions where activity is comparatively low, such as regions distal to the somatosensory cortex. Additionally, low SNR may result from electrodes being positioned directly above a sulcus, where the electrode may not detect the electric fields generated by pyramidal neurons due to their orientation. In either case, SNR variance between channels suggests the 1~mm and 3~mm electrodes record with spatial resolution high enough to discriminate between activity from neural populations in close proximity. 

\subsubsection{Correlation}
The channels appear to form distinct functional groups. Channels within each group are highly correlated. Conversely, channels from different groups exhibit strong negative correlations, indicating that they may be recording activity from the same source, but with an opposing polarity. Previous work has demonstrated that variation in phase can relate to the electrodes location relative to the SEP, particularly when the source is located in a sulcus \citep{schreiner_mapping_2024, schramm_localization_1991}. This is commonly seen in intracranial electrodes, such as ECoG \citep{vakani_electrocorticography_2019}, but has also been demonstrated non-invasively with ultra high density EEG \citep{schreiner_mapping_2024}. Most channels were either highly positively or negatively correlated. The observed differences in inter-channel correlation, and SNR, across channels are likely attributable to variations in the underlying brain anatomy, including sulcal and gyral patterns, skull thickness, and vasculature. These findings indicate that the sub-scalp EEG recordings have sufficient spatial resolution to differentiate neural activity originating from these various structures. However, the precise location of such structures relative to the arrays was not verified.

\subsubsection{Spatial Resolution}
Spatial resolution increased when using higher density arrays. 1~mm and 3~mm diameter electrodes with 5~mm pitch recorded activity with distinct variations in SNR and correlation between adjacent channels. Together, these results suggests 5~mm inter-channel distance is not spatially oversampling. Current sub-scalp EEG devices have been developed for chronic seizure monitoring or detection of hypoglycaemia \citep{duun-henriksen_new_2020, haneef_sub-scalp_2022}, typically featuring less than 8 channels and low sampling rates (250 Hz). Our findings indicate that higher density sub-scalp electrode arrays may have utility for brain-computer interface applications. Future work should aim to identify the minimum inter-channel distance for electrodes in the sub-scalp space through use of higher density arrays. 

\subsection{Aim 2: Sub-scalp EEG Motor Decoding}
The behavioural experiment results suggest that sub-scalp EEG is suitable for BCI applications that utilise motor-related neural activity. We discriminated between left and right movement execution in two of four sheep, and movement in either the left or right direction vs. rest in three of  four sheep. The results are comparable to similar work previously performed to test efficacy of endovascular electrode arrays for BCI applications \citep{forsyth_evaluation_2019}. Additionally, feature selection provided valuable insight into the importance of different temporal, spectral, and spatial qualities of sub-scalp EEG recordings.

\subsubsection{Feature Importance}
The features that were most important for movement classification were different across the animals, demonstrating the importance of animal (or human) specific models for feature selection and classification. Regarding the frequency space, high gamma features most often demonstrated high mutual information with the class labels. High gamma is a useful feature for BCI applications as it has been shown to have more functional localisation than lower frequency bands, although it is difficult to detect with non-invasive EEG due to low amplitudes \citep{schalk_brain-computer_2011, wolpaw_braincomputer_2012-1}. This is the first demonstration of detection of high gamma features associated with motor function with sub-scalp EEG. This result supports sub-scalp EEG use in BCI applications and warrants further investigation as part of an online BCI system. 

In contrast with prior understanding of sensorimotor activity in humans \citep{pfurtscheller_chapter_2000}, the beta band was of minimal importance, being the least selected band across all sheep and classification cases. This may be due to physiological differences between human and sheep models, or it is possible that the availability of high-quality gamma and high gamma features may have taken precedence during feature selection.



Spatially, we observed some variation between the median spectrograms of each channel (Figure~\ref{fig:CH4_Fig4_e}). Typically in human models, there is a strong contralateral relationship between power modulation in the motor cortex and movement that aids in discriminating between motor activity from the left or right side of the body \citep{pfurtscheller_chapter_2000}. In sheep models, the motor cortices of both hemispheres are proximal to each other and extend along the rostal-caudal axis. As such, potentials measured by electrodes outside the skull will likely contain a summation of activity from sources across both hemispheres. This physiology may explain the poor sensitivity and specificity of the left and right classes as spatial features are critical for this discrimination. Recording neural activity with high resolution is beneficial for BCI applications as discriminating between activity in proximal brain regions can facilitate higher degrees of freedom. We expect brain structure to present fewer challenges in humans than sheep models due to the larger size and lateral structure of the motor cortex. Future investigation focused on spatial resolution of sub-scalp electrodes would provide further insight.

\subsubsection{Classification}
Using linear discriminant analysis we discriminated motor-related neural activity significantly above chance. This alone would allow for simple binary outputs that are useful for certain BCI applications in persons with severe disability, such as alerting carers, clicking in digital environments that use an automated scroll, and binary communication protocols. Chronic sub-scalp EEG implantation can provide the opportunity to collect large datasets that allow for training of feature selection and machine learning algorithms with more parameters that may produce better performance and increased dimensional control for users. Previous studies have demonstrated two- \citep{dandan_huang_eeg-based_2009, kayagil_binary_2009} and three-dimensional control \citep{mcfarland_electroencephalographic_2010} using non-invasive EEG. We expect sub-scalp EEG BCI performance to at least match non-invasive counterparts with equivalent channel setups due to improved stability, and increased amplitude from avoiding attenuation through scalp tissue.

Demonstrating classification of motor cortex activity, despite the various challenges associated with the data collection for this experiment, support the use of sub-scalp BCI for in-home, everyday use. During data recording, the sheep were in noisy environments: surrounded by distractions of other noisy sheep, impatient for food, chewing, and vocalising. There was no Faraday cage to remove external electrical interference. Additionally, while the animals did perform left and right head movements, these movements were often inconsistent, paired with upward or downward motion, slow or fast, and preceded by inactivity, eating, vocalising, stomping, or other motions. All of these factors contributed to EEG noise in this experiment and, despite these factors, we were able to discriminate motor activity. Unlike controlled laboratory or clinical environments, in-home BCI users will be surrounded by other devices, people, and distractions, and will not utilise large, expensive, sophisticated amplification equipment. Reliable control is essential for user acceptance of BCI devices. This study has demonstrated sub-scalp EEG has the potential to provide reliability as a sensorimotor BCI outside of controlled laboratory environments.

In this experiment, sub-scalp EEG has demonstrated comparable performance with similar work investigating endovascular electrode arrays. \citet{forsyth_evaluation_2019} were able to decode left vs. right discrimination significantly above chance in 50\% of sheep with endovascular data (maximum accuracy = 70\%, n=2), as were we with sub-scalp electrodes (test set accuracy = 66\%, n=4). Discrimination between grouped left and right movement and rest was achieved in both sheep with endovascular electrodes, reporting a maximum classification accuracy of 80\% and 90\% \citep{forsyth_evaluation_2019}. While we did not group both directions into a single class, similar performance was observed for rest class sensitivity (74±1\%) and specificity (82±5\%). The comparable signal quality between sub-scalp EEG in this study and endovascular electrodes as per \citet{forsyth_evaluation_2019} is consistent with previous work comparing these two recording modalities \citep{mahoney_comparison_2023}. Since the work of \citet{forsyth_evaluation_2019}, the endovascular BCI approach has progressed into human trials, demonstrating efficacy as a chronic, in-home system by providing benefit to persons with ALS \citep{mitchell_assessment_2023}. 

While comparable motor decoding studies in sheep models are limited, substantial literature exists regarding motor decoding in humans across non-invasive EEG, ECoG, and endovascular modalities. ECoG arrays have consistently achieved high-accuracy decoding for a broad spectrum of motor activity, including hand and elbow postures \citep{yanagisawa_electrocorticographic_2012}, finger movement \citep{yao_fast_2022}, and speech \citep{littlejohn_streaming_2025}. A direct comparison by \citet{liao_decoding_2014} on pairwise finger movement decoding between non-invasive EEG and ECoG revealed average accuracies of 77\% and 91\%, respectively. Similarly, studies on speech decoding from non-invasive EEG have demonstrated high classification accuracy ($>$80\%) for limited vocabularies (e.g., ``left", ``down", ``right", ``up", and ``select") \citep{moctezuma_subjects_2019, wang_eeg-based_2018}. However, these have not yet achieved the accuracy levels comparable to ECoG speech decoding systems \citep{lopez-bernal_state---art_2022}. ECoG recordings provide a clear improvement in BCI functionality over non-invasive EEG. While signal quality of endovascular electrodes has been shown to be generally higher than non-invasive EEG \citep{thielen_making_2023} and comparable to ECoG \citep{john_signal_2018}, this fidelity has not yet translated to BCI functionality. Preliminary in-human endovascular BCI studies have reported classification accuracies of 91\% and 70\% (from two patients) for four gross hand and foot motor execution classes \citep{kacker_motor_2025}, providing a benchmark for current minimally invasive BCI performance. However, as endovascular arrays evolve to access smaller, peripheral vasculature and record from broader neural populations, improved BCI performance is anticipated. Given the inherent challenges of replicating complex motor decoding experiments in animal models, future investigations should prioritise evaluating the performance of sub-scalp EEG as a BCI in human subjects, to determine if this modality can achieve comparable success to that demonstrated here in sheep models.

\subsection{Feasibility of Sub-scalp BCI}
\label{limiations}
Overall, we have completed both aims of this study and provide evidence in support of sub-scalp EEG for BCI applications. BCIs have demonstrated functionality across diverse signal acquisition modalities, including non-invasive EEG and invasive techniques like ECoG and endovascular electrodes. The spatial resolution of the signal acquisition methods significantly impacts the capabilities of the BCI system. High EEG spatial resolution means the recordings can be attributed to highly localised regions of the brain. Humans have demonstrated the ability to learn to modulate regions of the motor cortex when provided with feedback \citep{mcfarland_electroencephalographic_2010, wolpaw_control_2004, mcfarland_eeg-based_2017}. In the context of BCI applications, decoded modulations can provide input into any number of devices, allowing the user direct control of the device through their neural activity. High spatial resolution means the ability to decode from more localised regions of the cortex. If the user learns to modulate these localised regions independently, this leads to higher dimensional control that, in turn, results in improved BCI function and user acceptance.

While spatial resolution of EEG is typically stated as in the order of centimetres \citep{slutzky_optimal_2010, robert_spitzer_method_1989, srinivasan_spatial_1998, freeman_spatial_2003}, higher density, non-invasive arrays have been shown to provide high signal quality, and even identify underlying gyri and sulci \citep{schreiner_mapping_2024}. High density arrays can isolate activity from local brain regions that may lead to high functioning BCIs \citep{lee_individual_2022}. However, as the number of electrodes increases, so too does the difficulty of setting up and maintaining non-invasive EEG systems. Incorporating high-density arrays into sub-scalp EEG systems results in no added setup or maintenance from the user’s perspective. In this way, sub-scalp EEG systems offer a continuous, long-term recording solution capable of signal quality equivalent to the highest resolution non-invasive EEG systems currently available, and with none of the daily setup.

ECoG and endovascular electrodes have demonstrated an optimal inter-channel distance between 1 and 10~mm in humans \citep{slutzky_optimal_2010, freeman_spatial_2000} and sheep models \citep{john_signal_2018}. The results of this study suggest that the optimal inter-channel distance for sub-scalp EEG recordings is 5~mm or less, indicating that sub-scalp EEG spatial resolution may be closer to intracranial than non-invasive electrodes. While these results are preliminary and subject to limitations, they offer evidence that sub-scalp EEG can record with high spatial resolution, and may allow for high performance BCIs without the need to enter the skull. ECoG BCI systems have demonstrated high bandwidth and utility by allowing users to communicate with BCI spellers, or control computers, prosthetics, and wheelchairs \citep{miller_current_2020}. More recently, ECoG arrays has shown promise as part of BCIs for decoding non-verbal user's speech  \citep{metzger_generalizable_2022, silva_speech_2024,littlejohn_streaming_2025}. These technologies may be of assistance to a broad range of cohorts with physical and verbal disability. However, the risk of sub-cranial implantation may deter these cohorts from accessing ECoG BCIs. Endovascular arrays mitigate this risk, however are still more invasive than sub-scalp arrays, and carry risk of stenosis and thrombosis \cite{starke_endovascular_2015, modi_stent_2024, soldozy_systematic_2020}. We demonstrated comparable motor decoding performance between endovascular and sub-scalp arrays. Should sub-scalp BCI demonstrate comparable utility to ECoG BCIs, it has the potential for tremendous impact, providing significant improvement to quality of life by restoring fundamental abilities such as speech and mobility, across a broad range of users.

The major limitation of this work is the use of sheep models. Sheep models have significant anatomical differences to human models, including skull thickness, brain size, and sensory and motor cortex structure. While we have also included and compared with results from previous studies that have explored ECoG and endovascular spatial resolution in sheep models \citep{john_signal_2018}, future work should endeavour to determine sub-scalp EEG signal qualities and BCI performance in humans.

\section{Conclusion}
Non-invasive recording of sensorimotor neural activity has demonstrated utility as input for BCI applications in previous studies. However, user acceptance of these systems for chronic BCI applications is low. Sub-scalp EEG is a minimally invasive alternative BCI signal acquisition method that addresses limitations of existing and emerging chronic technologies. We have demonstrated sub-scalp recording of sensorimotor rhythms in sheep models, identified important spatial, temporal, and spectral features, and classified motor execution with above chance performance. Our results are comparable to previous work investigating SEP signal quality \citep{john_signal_2018} and classification of motor-related neural activity \citep{forsyth_evaluation_2019} using ECoG and endovascular arrays in sheep models. These findings support further investigation into sub-scalp BCI technology as a tool for assisting persons with chronic physical disability with everyday activities. Future work should focus on characterising sub-scalp EEG spatial resolution, understanding the potential BCI user acceptance of sub-scalp technology, and in-human sub-scalp BCI clinical studies.

\ack
We thank veterinary technician Tomas Vale and animal handler Quan Nguyen (The Florey Institute of Neuroscience and Mental Health) for their assistance in caring for the animals, surgical preparations, and monitoring during the animal experiments.

We also express our appreciation for the animals used for this research. Animal studies are crucial for the realisation of medical devices. We intend to ensure insights gained from their sacrifice lead to improved quality of life for persons with severe disability.

TB Mahoney was supported by the Australian Government Research Training Program Scholarship, Elizabeth and Vernon Puzey Scholarship, and the PhD Write-Up Award (Faculty of Engineering and Information Technology) from the University of Melbourne, Australia.

There are no conflicts of interest to report.

All raw data is available through the Figshare database for experiments pertaining to both Aim 1 \citep{mahoney_sep_2025} and Aim 2 \citep{mahoney_behavioural_2025}.


\begin{thebibliography}{76}
\providecommand{\natexlab}[1]{#1}
\providecommand{\url}[1]{\texttt{#1}}
\expandafter\ifx\csname urlstyle\endcsname\relax
  \providecommand{\doi}[1]{doi: #1}\else
  \providecommand{\doi}{doi: \begingroup \urlstyle{rm}\Url}\fi

\bibitem[Zedde et~al.(2022)Zedde, Grisendi, Pezzella, Napoli, Moratti, Valzania, and Pascarella]{zedde_acute_2022}
Marialuisa Zedde, Ilaria Grisendi, Francesca~Romana Pezzella, Manuela Napoli, Claudio Moratti, Franco Valzania, and Rosario Pascarella.
\newblock Acute {Onset} {Quadriplegia} and {Stroke}: {Look} at the {Brainstem}, {Look} at the {Midline}.
\newblock \emph{Journal of Clinical Medicine}, 11\penalty0 (23):\penalty0 7205, December 2022.
\newblock ISSN 2077-0383.
\newblock \doi{10.3390/jcm11237205}.
\newblock URL \url{https://www.mdpi.com/2077-0383/11/23/7205}.

\bibitem[Haberl et~al.(1994)Haberl, Vollmer, and Hacke]{hacke_tetraplegia_1994}
Roman Haberl, Dennis~G. Vollmer, and Werner Hacke.
\newblock Tetraplegia and {Paraplegia}.
\newblock In Werner Hacke, Daniel~F. Hanley, Karl~M. Einhäupl, Thomas~P. Bleck, Michael~N. Diringer, and Allan~H. Ropper, editors, \emph{Neurocritical {Care}}, pages 292--306. Springer Berlin Heidelberg, Berlin, Heidelberg, 1994.
\newblock ISBN 978-3-642-87604-2 978-3-642-87602-8.
\newblock \doi{10.1007/978-3-642-87602-8_28}.
\newblock URL \url{http://link.springer.com/10.1007/978-3-642-87602-8_28}.

\bibitem[Sunny et~al.(2021)Sunny, Zarif, Rulik, Sanjuan, Rahman, Ahamed, Wang, Schultz, and Brahmi]{sunny_eye-gaze_2021}
Md~Samiul~Haque Sunny, Md~Ishrak~Islam Zarif, Ivan Rulik, Javier Sanjuan, Mohammad~Habibur Rahman, Sheikh~Iqbal Ahamed, Inga Wang, Katie Schultz, and Brahim Brahmi.
\newblock Eye-gaze control of a wheelchair mounted {6DOF} assistive robot for activities of daily living.
\newblock \emph{Journal of NeuroEngineering and Rehabilitation}, 18\penalty0 (1):\penalty0 173, December 2021.
\newblock ISSN 1743-0003.
\newblock \doi{10.1186/s12984-021-00969-2}.
\newblock URL \url{https://jneuroengrehab.biomedcentral.com/articles/10.1186/s12984-021-00969-2}.

\bibitem[Bissoli et~al.(2019)Bissoli, Lavino-Junior, Sime, Encarnação, and Bastos-Filho]{bissoli_humanmachine_2019}
Alexandre Bissoli, Daniel Lavino-Junior, Mariana Sime, Lucas Encarnação, and Teodiano Bastos-Filho.
\newblock A {Human}–{Machine} {Interface} {Based} on {Eye} {Tracking} for {Controlling} and {Monitoring} a {Smart} {Home} {Using} the {Internet} of {Things}.
\newblock \emph{Sensors}, 19\penalty0 (4):\penalty0 859, February 2019.
\newblock ISSN 1424-8220.
\newblock \doi{10.3390/s19040859}.
\newblock URL \url{http://www.mdpi.com/1424-8220/19/4/859}.

\bibitem[Kübler et~al.(2005)Kübler, Nijboer, Mellinger, Vaughan, Pawelzik, Schalk, McFarland, Birbaumer, and Wolpaw]{kubler_patients_2005}
A.~Kübler, F.~Nijboer, J.~Mellinger, T.~M. Vaughan, H.~Pawelzik, G.~Schalk, D.~J. McFarland, N.~Birbaumer, and J.~R. Wolpaw.
\newblock Patients with {ALS} can use sensorimotor rhythms to operate a brain-computer interface.
\newblock \emph{Neurology}, 64\penalty0 (10):\penalty0 1775--7, 2005.
\newblock ISSN 0028-3878.
\newblock \doi{10.1212/01.Wnl.0000158616.43002.6d}.
\newblock Type: Journal Article.

\bibitem[{Han Yuan} and {Bin He}(2014)]{han_yuan_braincomputer_2014}
{Han Yuan} and {Bin He}.
\newblock Brain–{Computer} {Interfaces} {Using} {Sensorimotor} {Rhythms}: {Current} {State} and {Future} {Perspectives}.
\newblock \emph{IEEE Transactions on Biomedical Engineering}, 61\penalty0 (5):\penalty0 1425--1435, May 2014.
\newblock ISSN 0018-9294, 1558-2531.
\newblock \doi{10.1109/TBME.2014.2312397}.
\newblock URL \url{http://ieeexplore.ieee.org/document/6775293/}.

\bibitem[Peters et~al.(2015)Peters, Bieker, Heckman, Huggins, Wolf, Zeitlin, and Fried-Oken]{peters_brain-computer_2015}
Betts Peters, Gregory Bieker, Susan~M. Heckman, Jane~E. Huggins, Catherine Wolf, Debra Zeitlin, and Melanie Fried-Oken.
\newblock Brain-{Computer} {Interface} {Users} {Speak} {Up}: {The} {Virtual} {Users}' {Forum} at the 2013 {International} {Brain}-{Computer} {Interface} {Meeting}.
\newblock \emph{Archives of Physical Medicine and Rehabilitation}, 96\penalty0 (3, Supplement):\penalty0 S33--S37, 2015.
\newblock ISSN 0003-9993.
\newblock \doi{10.1016/j.apmr.2014.03.037}.
\newblock Type: Journal Article.

\bibitem[Blabe et~al.(2015)Blabe, Gilja, Chestek, Shenoy, Anderson, and Henderson]{blabe_assessment_2015}
Christine~H. Blabe, Vikash Gilja, Cindy~A. Chestek, Krishna~V. Shenoy, Kim~D. Anderson, and Jaimie~M. Henderson.
\newblock Assessment of brain-machine interfaces from the perspective of people with paralysis.
\newblock \emph{Journal of Neural Engineering}, 12\penalty0 (4):\penalty0 043002, 2015.
\newblock ISSN 1741-2560 1741-2552.
\newblock \doi{10.1088/1741-2560/12/4/043002}.
\newblock Type: Journal Article.

\bibitem[Kubler et~al.(2014)Kubler, Holz, Riccio, Zickler, Kaufmann, Kleih, Staiger-Salzer, Desideri, Hoogerwerf, and Mattia]{kubler_user-centered_2014}
A.~Kubler, E.~M. Holz, A.~Riccio, C.~Zickler, T.~Kaufmann, S.~C. Kleih, P.~Staiger-Salzer, L.~Desideri, E.~J. Hoogerwerf, and D.~Mattia.
\newblock The user-centered design as novel perspective for evaluating the usability of {BCI}-controlled applications.
\newblock \emph{PLoS One}, 9\penalty0 (12):\penalty0 e112392, 2014.
\newblock ISSN 1932-6203 (Electronic) 1932-6203 (Linking).
\newblock \doi{10.1371/journal.pone.0112392}.
\newblock Type: Journal Article.

\bibitem[Käthner et~al.(2017)Käthner, Halder, Hintermüller, Espinosa, Guger, Miralles, Vargiu, Dauwalder, Rafael-Palou, Solà, Daly, Armstrong, Martin, and Kübler]{kathner_multifunctional_2017}
Ivo Käthner, Sebastian Halder, Christoph Hintermüller, Arnau Espinosa, Christoph Guger, Felip Miralles, Eloisa Vargiu, Stefan Dauwalder, Xavier Rafael-Palou, Marc Solà, Jean~M. Daly, Elaine Armstrong, Suzanne Martin, and Andrea Kübler.
\newblock A {Multifunctional} {Brain}-{Computer} {Interface} {Intended} for {Home} {Use}: {An} {Evaluation} with {Healthy} {Participants} and {Potential} {End} {Users} with {Dry} and {Gel}-{Based} {Electrodes}.
\newblock \emph{Frontiers in Neuroscience}, 11:\penalty0 286, May 2017.
\newblock ISSN 1662-453X.
\newblock \doi{10.3389/fnins.2017.00286}.
\newblock URL \url{http://journal.frontiersin.org/article/10.3389/fnins.2017.00286/full}.

\bibitem[Cardoso et~al.(2022)Cardoso, Andreasen~Struijk, Kaeseler, and Jochumsen]{cardoso_comparing_2022}
Ana S.~S. Cardoso, Lotte N.~S. Andreasen~Struijk, Rasmus~L. Kaeseler, and Mads Jochumsen.
\newblock Comparing the {Usability} of {Alternative} {EEG} {Devices} to {Traditional} {Electrode} {Caps} for {SSVEP}-{BCI} {Controlled} {Assistive} {Robots}.
\newblock In \emph{2022 {International} {Conference} on {Rehabilitation} {Robotics} ({ICORR})}, pages 1--6, Rotterdam, Netherlands, July 2022. IEEE.
\newblock ISBN 9781665488297.
\newblock \doi{10.1109/ICORR55369.2022.9896588}.
\newblock URL \url{https://ieeexplore.ieee.org/document/9896588/}.

\bibitem[Brannigan et~al.(2024{\natexlab{a}})Brannigan, Liyanage, Horsfall, Bashford, Muirhead, and Fry]{brannigan_braincomputer_2024}
Jamie F~M Brannigan, Kishan Liyanage, Hugo~Layard Horsfall, Luke Bashford, William Muirhead, and Adam Fry.
\newblock Brain–computer interfaces patient preferences: a systematic review.
\newblock \emph{Journal of Neural Engineering}, 21\penalty0 (6):\penalty0 061005, December 2024{\natexlab{a}}.
\newblock ISSN 1741-2560, 1741-2552.
\newblock \doi{10.1088/1741-2552/ad94a6}.
\newblock URL \url{https://iopscience.iop.org/article/10.1088/1741-2552/ad94a6}.

\bibitem[Oxley et~al.(2021)Oxley, Yoo, Rind, Ronayne, Lee, Bird, Hampshire, Sharma, Morokoff, Williams, MacIsaac, Howard, Irving, Vrljic, Williams, John, Weissenborn, Dazenko, Balabanski, Friedenberg, Burkitt, Wong, Drummond, Desmond, Weber, Denison, Hochberg, Mathers, O’Brien, May, Mocco, Grayden, Campbell, Mitchell, and Opie]{oxley_motor_2021}
Thomas~J Oxley, Peter~E Yoo, Gil~S Rind, Stephen~M Ronayne, C~M~Sarah Lee, Christin Bird, Victoria Hampshire, Rahul~P Sharma, Andrew Morokoff, Daryl~L Williams, Christopher MacIsaac, Mark~E Howard, Lou Irving, Ivan Vrljic, Cameron Williams, Sam~E John, Frank Weissenborn, Madeleine Dazenko, Anna~H Balabanski, David Friedenberg, Anthony~N Burkitt, Yan~T Wong, Katharine~J Drummond, Patricia Desmond, Douglas Weber, Timothy Denison, Leigh~R Hochberg, Susan Mathers, Terence~J O’Brien, Clive~N May, J~Mocco, David~B Grayden, Bruce C~V Campbell, Peter Mitchell, and Nicholas~L Opie.
\newblock Motor neuroprosthesis implanted with neurointerventional surgery improves capacity for activities of daily living tasks in severe paralysis: first in-human experience.
\newblock \emph{Journal of NeuroInterventional Surgery}, 13\penalty0 (2):\penalty0 102--108, 2021.
\newblock ISSN 1759-8478.
\newblock \doi{10.1136/neurintsurg-2020-016862}.
\newblock URL \url{https://jnis.bmj.com/content/13/2/102}.

\bibitem[Brannigan et~al.(2024{\natexlab{b}})Brannigan, Fry, Opie, Campbell, Mitchell, and Oxley]{brannigan_endovascular_2024}
Jamie~F.M. Brannigan, Adam Fry, Nicholas~L. Opie, Bruce~C.V. Campbell, Peter~J. Mitchell, and Thomas~J. Oxley.
\newblock Endovascular {Brain}-{Computer} {Interfaces} in {Poststroke} {Paralysis}.
\newblock \emph{Stroke}, 55\penalty0 (2):\penalty0 474--483, February 2024{\natexlab{b}}.
\newblock ISSN 0039-2499, 1524-4628.
\newblock \doi{10.1161/STROKEAHA.123.037719}.
\newblock URL \url{https://www.ahajournals.org/doi/10.1161/STROKEAHA.123.037719}.

\bibitem[Zhang et~al.(2023)Zhang, Mandeville, Xu, Stary, Lo, and Lieber]{zhang_ultraflexible_2023}
Anqi Zhang, Emiri~T. Mandeville, Lijun Xu, Creed~M. Stary, Eng~H. Lo, and Charles~M. Lieber.
\newblock Ultraflexible endovascular probes for brain recording through micrometer-scale vasculature.
\newblock \emph{Science}, 381\penalty0 (6655):\penalty0 306--312, July 2023.
\newblock ISSN 0036-8075, 1095-9203.
\newblock \doi{10.1126/science.adh3916}.
\newblock URL \url{https://www.science.org/doi/10.1126/science.adh3916}.

\bibitem[Starke et~al.(2015)Starke, Wang, Ding, Durst, Crowley, Chalouhi, Hasan, Dumont, Jabbour, and Liu]{starke_endovascular_2015}
Robert~M. Starke, Tony Wang, Dale Ding, Christopher~R. Durst, R.~Webster Crowley, Nohra Chalouhi, David~M. Hasan, Aaron~S. Dumont, Pascal Jabbour, and Kenneth~C. Liu.
\newblock Endovascular {Treatment} of {Venous} {Sinus} {Stenosis} in {Idiopathic} {Intracranial} {Hypertension}: {Complications}, {Neurological} {Outcomes}, and {Radiographic} {Results}.
\newblock \emph{The Scientific World Journal}, 2015:\penalty0 140408, 2015.
\newblock ISSN 2356-6140.
\newblock \doi{10.1155/2015/140408}.
\newblock URL \url{https://doi.org/10.1155/2015/140408}.
\newblock Type: Journal Article.

\bibitem[Modi et~al.(2024)Modi, Soos, and Mahajan]{modi_stent_2024}
Kalgi Modi, Michael~P. Soos, and Kunal Mahajan.
\newblock Stent {Thrombosis}.
\newblock In \emph{{StatPearls}}. StatPearls Publishing, Treasure Island (FL), 2024.
\newblock URL \url{http://www.ncbi.nlm.nih.gov/books/NBK441908/}.

\bibitem[Soldozy et~al.(2020)Soldozy, Young, Kumar, Capek, Felbaum, Jean, Park, and Syed]{soldozy_systematic_2020}
Sauson Soldozy, Steven Young, Jeyan~S. Kumar, Stepan Capek, Daniel~R. Felbaum, Walter~C. Jean, Min~S. Park, and Hasan~R. Syed.
\newblock A systematic review of endovascular stent-electrode arrays, a minimally invasive approach to brain-machine interfaces.
\newblock \emph{Neurosurgical Focus}, 49\penalty0 (1):\penalty0 E3, July 2020.
\newblock ISSN 1092-0684.
\newblock \doi{10.3171/2020.4.FOCUS20186}.
\newblock URL \url{https://thejns.org/view/journals/neurosurg-focus/49/1/article-pE3.xml}.

\bibitem[Stirling et~al.(2021)Stirling, Maturana, Karoly, Nurse, McCutcheon, Grayden, Ringo, Heasman, Hoare, Lai, D'Souza, Seneviratne, Seiderer, McLean, Bulluss, Murphy, Brinkmann, Richardson, Freestone, and Cook]{stirling_seizure_2021}
Rachel~E. Stirling, Matias~I. Maturana, Philippa~J. Karoly, Ewan~S. Nurse, Kate McCutcheon, David~B. Grayden, Steven~G. Ringo, John~M. Heasman, Rohan~J. Hoare, Alan Lai, Wendyl D'Souza, Udaya Seneviratne, Linda Seiderer, Karen~J. McLean, Kristian~J. Bulluss, Michael Murphy, Benjamin~H. Brinkmann, Mark~P. Richardson, Dean~R. Freestone, and Mark~J. Cook.
\newblock Seizure {Forecasting} {Using} a {Novel} {Sub}-{Scalp} {Ultra}-{Long} {Term} {EEG} {Monitoring} {System}.
\newblock \emph{Frontiers in Neurology}, 12\penalty0 (1445), 2021.
\newblock ISSN 1664-2295.
\newblock \doi{10.3389/fneur.2021.713794}.
\newblock Type: Journal Article.

\bibitem[Duun-Henriksen et~al.(2020)Duun-Henriksen, Baud, Richardson, Cook, Kouvas, Heasman, Friedman, Peltola, Zibrandtsen, and Kjaer]{duun-henriksen_new_2020}
Jonas Duun-Henriksen, Maxime Baud, Mark~P. Richardson, Mark Cook, George Kouvas, John~M. Heasman, Daniel Friedman, Jukka Peltola, Ivan~C. Zibrandtsen, and Troels~W. Kjaer.
\newblock A new era in electroencephalographic monitoring? {Subscalp} devices for ultra-long-term recordings.
\newblock \emph{EPILEPSIA}, 2020.
\newblock ISSN 00139580.
\newblock \doi{10.1111/epi.16630}.
\newblock Type: Journal Article.

\bibitem[Weisdorf et~al.(2019)Weisdorf, Duun-Henriksen, Kjeldsen, Poulsen, Gangstad, and Kjær]{weisdorf_ultra-long-term_2019}
Sigge Weisdorf, Jonas Duun-Henriksen, Marianne~J. Kjeldsen, Frantz~R. Poulsen, Sirin~W. Gangstad, and Troels~W. Kjær.
\newblock Ultra-long-term subcutaneous home monitoring of epilepsy—490 days of {EEG} from nine patients.
\newblock \emph{Epilepsia}, 60\penalty0 (11):\penalty0 2204--2214, 2019.
\newblock ISSN 0013-9580.
\newblock \doi{10.1111/epi.16360}.
\newblock Type: Journal Article.

\bibitem[Barlatey et~al.(2024)Barlatey, Kouvas, Sobolewski, Nowacki, Pollo, and Baud]{barlatey_designing_2024}
Sabry~L. Barlatey, George Kouvas, Aleksander Sobolewski, Andreas Nowacki, Claudio Pollo, and Maxime~O. Baud.
\newblock Designing next-generation subscalp devices for seizure monitoring: {A} systematic review and meta-analysis of established extracranial hardware.
\newblock \emph{Epilepsy Research}, 202:\penalty0 107356, May 2024.
\newblock ISSN 09201211.
\newblock \doi{10.1016/j.eplepsyres.2024.107356}.
\newblock URL \url{https://linkinghub.elsevier.com/retrieve/pii/S0920121124000718}.

\bibitem[Haneef et~al.(2022)Haneef, Yang, Sheth, Aloor, Aazhang, Krishnan, and Karakas]{haneef_sub-scalp_2022}
Zulfi Haneef, Kaiyuan Yang, Sameer~A. Sheth, Fuad~Z. Aloor, Behnaam Aazhang, Vaishnav Krishnan, and Cemal Karakas.
\newblock Sub-scalp electroencephalography: {A} next-generation technique to study human neurophysiology.
\newblock \emph{Clinical Neurophysiology}, 141:\penalty0 77--87, 2022.
\newblock ISSN 1388-2457.
\newblock \doi{10.1016/j.clinph.2022.07.003}.
\newblock URL \url{https://www.sciencedirect.com/science/article/pii/S1388245722003273}.

\bibitem[{Dandan Huang} et~al.(2009){Dandan Huang}, Lin, {Ding-Yu Fei}, {Xuedong Chen}, and {Ou Bai}]{dandan_huang_eeg-based_2009}
{Dandan Huang}, P.~Lin, {Ding-Yu Fei}, {Xuedong Chen}, and {Ou Bai}.
\newblock {EEG}-based online two-dimensional cursor control.
\newblock In \emph{2009 {Annual} {International} {Conference} of the {IEEE} {Engineering} in {Medicine} and {Biology} {Society}}, pages 4547--4550, Minneapolis, MN, September 2009. IEEE.
\newblock \doi{10.1109/IEMBS.2009.5332722}.
\newblock URL \url{http://ieeexplore.ieee.org/document/5332722/}.

\bibitem[Kayagil et~al.(2009)Kayagil, Bai, Henriquez, Lin, Furlani, Vorbach, and Hallett]{kayagil_binary_2009}
Turan~A Kayagil, Ou~Bai, Craig~S Henriquez, Peter Lin, Stephen~J Furlani, Sherry Vorbach, and Mark Hallett.
\newblock A binary method for simple and accurate two-dimensional cursor control from {EEG} with minimal subject training.
\newblock \emph{Journal of NeuroEngineering and Rehabilitation}, 6\penalty0 (1):\penalty0 14, December 2009.
\newblock ISSN 1743-0003.
\newblock \doi{10.1186/1743-0003-6-14}.
\newblock URL \url{https://jneuroengrehab.biomedcentral.com/articles/10.1186/1743-0003-6-14}.

\bibitem[McFarland et~al.(2010)McFarland, Sarnacki, and Wolpaw]{mcfarland_electroencephalographic_2010}
Dennis~J McFarland, William~A Sarnacki, and Jonathan~R Wolpaw.
\newblock Electroencephalographic ({EEG}) control of three-dimensional movement.
\newblock \emph{Journal of Neural Engineering}, 7\penalty0 (3):\penalty0 036007, June 2010.
\newblock ISSN 1741-2560, 1741-2552.
\newblock \doi{10.1088/1741-2560/7/3/036007}.
\newblock URL \url{https://iopscience.iop.org/article/10.1088/1741-2560/7/3/036007}.

\bibitem[Littlejohn et~al.(2025)Littlejohn, Cho, Liu, Silva, Yu, Anderson, Kurtz-Miott, Brosler, Kashyap, Hallinan, Shah, Tu-Chan, Ganguly, Moses, Chang, and Anumanchipalli]{littlejohn_streaming_2025}
Kaylo~T. Littlejohn, Cheol~Jun Cho, Jessie~R. Liu, Alexander~B. Silva, Bohan Yu, Vanessa~R. Anderson, Cady~M. Kurtz-Miott, Samantha Brosler, Anshul~P. Kashyap, Irina~P. Hallinan, Adit Shah, Adelyn Tu-Chan, Karunesh Ganguly, David~A. Moses, Edward~F. Chang, and Gopala~K. Anumanchipalli.
\newblock A streaming brain-to-voice neuroprosthesis to restore naturalistic communication.
\newblock \emph{Nature Neuroscience}, 28\penalty0 (4):\penalty0 902--912, April 2025.
\newblock ISSN 1097-6256, 1546-1726.
\newblock \doi{10.1038/s41593-025-01905-6}.
\newblock URL \url{https://www.nature.com/articles/s41593-025-01905-6}.

\bibitem[Nunez and Srinivasan(2006)]{nunez_electric_2006}
Paul~L. Nunez and Ramesh Srinivasan.
\newblock \emph{Electric {Fields} of the {Brain}}.
\newblock Oxford University Press, January 2006.
\newblock ISBN 978-0-19-505038-7.
\newblock \doi{10.1093/acprof:oso/9780195050387.001.0001}.
\newblock URL \url{http://www.oxfordscholarship.com/view/10.1093/acprof:oso/9780195050387.001.0001/acprof-9780195050387}.

\bibitem[Young et~al.(2006)Young, Ives, Chapman, and Mirsattari]{young_comparison_2006}
G.~Bryan Young, John~R. Ives, Martin~G. Chapman, and Seyed~M. Mirsattari.
\newblock A comparison of subdermal wire electrodes with collodion-applied disk electrodes in long-term {EEG} recordings in {ICU}.
\newblock \emph{Clinical Neurophysiology}, 117\penalty0 (6):\penalty0 1376--1379, 2006.
\newblock ISSN 1388-2457.
\newblock \doi{10.1016/j.clinph.2006.02.006}.
\newblock Type: Journal Article.

\bibitem[Duun-Henriksen et~al.(2015)Duun-Henriksen, Kjaer, Looney, Atkins, Sørensen, Rose, Mandic, Madsen, and Juhl]{duun-henriksen_eeg_2015}
Jonas Duun-Henriksen, Troels~Wesenberg Kjaer, David Looney, Mary~Doreen Atkins, Jens~Ahm Sørensen, Martin Rose, Danilo~P. Mandic, Rasmus~Elsborg Madsen, and Claus~Bogh Juhl.
\newblock {EEG} {Signal} {Quality} of a {Subcutaneous} {Recording} {System} {Compared} to {Standard} {Surface} {Electrodes}.
\newblock \emph{Journal of Sensors}, 2015:\penalty0 341208, 2015.
\newblock ISSN 1687-725X.
\newblock \doi{10.1155/2015/341208}.
\newblock Type: Journal Article.

\bibitem[Weisdorf et~al.(2018)Weisdorf, Gangstad, Duun-Henriksen, Mosholt, and Kjær]{weisdorf_high_2018}
Sigge Weisdorf, Sirin~W. Gangstad, Jonas Duun-Henriksen, Karina S.~S. Mosholt, and Troels~W. Kjær.
\newblock High similarity between {EEG} from subcutaneous and proximate scalp electrodes in patients with temporal lobe epilepsy.
\newblock \emph{Journal of Neurophysiology}, 120\penalty0 (3):\penalty0 1451--1460, 2018.
\newblock ISSN 0022-3077.
\newblock \doi{10.1152/jn.00320.2018}.
\newblock Type: Journal Article.

\bibitem[Olson et~al.(2016)Olson, Wander, Johnson, Sarma, Weaver, Novotny, Ojemann, and Darvas]{olson_comparison_2016}
Jared~D. Olson, Jeremiah~D. Wander, Lise Johnson, Devapratim Sarma, Kurt Weaver, Edward~J. Novotny, Jeffrey~G. Ojemann, and Felix Darvas.
\newblock Comparison of subdural and subgaleal recordings of cortical high-gamma activity in humans.
\newblock \emph{Clinical neurophysiology : official journal of the International Federation of Clinical Neurophysiology}, 127\penalty0 (1):\penalty0 277--284, 2016.
\newblock ISSN 1872-8952 1388-2457.
\newblock \doi{10.1016/j.clinph.2015.03.014}.
\newblock Type: Journal Article.

\bibitem[Mahoney et~al.(2023)Mahoney, Liu, Grayden, and John]{mahoney_comparison_2023}
Timothy~B. Mahoney, Po-Chen Liu, David~B. Grayden, and Sam~E. John.
\newblock Comparison of {Sub}-{Scalp} {EEG} and {Endovascular} {Stent}-{Electrode} {Array} for {Visual} {Evoked} {Potential} {Brain}-{Computer} {Interface}.
\newblock In \emph{2023 45th {Annual} {International} {Conference} of the {IEEE} {Engineering} in {Medicine} \& {Biology} {Society} ({EMBC})}, pages 1--4, Sydney, Australia, July 2023. IEEE.
\newblock ISBN 9798350324471.
\newblock \doi{10.1109/EMBC40787.2023.10340834}.
\newblock URL \url{https://ieeexplore.ieee.org/document/10340834/}.

\bibitem[Yang et~al.(2020)Yang, Tong, Shu, Zhuang, Yan, and Zeng]{yang_high_2020}
K.~Yang, L.~Tong, J.~Shu, N.~Zhuang, B.~Yan, and Y.~Zeng.
\newblock High {Gamma} {Band} {EEG} {Closely} {Related} to {Emotion}: {Evidence} {From} {Functional} {Network}.
\newblock \emph{Front Hum Neurosci}, 14:\penalty0 89, 2020.
\newblock ISSN 1662-5161 (Print) 1662-5161.
\newblock \doi{10.3389/fnhum.2020.00089}.
\newblock Type: Journal Article.

\bibitem[Fifer et~al.(2014)Fifer, Hotson, Wester, McMullen, Wang, Johannes, Katyal, Helder, Para, Vogelstein, Anderson, Thakor, and Crone]{fifer_simultaneous_2014}
Matthew~S. Fifer, Guy Hotson, Brock~A. Wester, David~P. McMullen, Yujing Wang, Matthew~S. Johannes, Kapil~D. Katyal, John~B. Helder, Matthew~P. Para, R.~Jacob Vogelstein, William~S. Anderson, Nitish~V. Thakor, and Nathan~E. Crone.
\newblock Simultaneous {Neural} {Control} of {Simple} {Reaching} and {Grasping} {With} the {Modular} {Prosthetic} {Limb} {Using} {Intracranial} {EEG}.
\newblock \emph{IEEE Transactions on Neural Systems and Rehabilitation Engineering}, 22\penalty0 (3):\penalty0 695--705, 2014.
\newblock \doi{10.1109/TNSRE.2013.2286955}.

\bibitem[Fitzgerald and Watson(2018)]{fitzgerald_gamma_2018}
Paul~J. Fitzgerald and Brendon~O. Watson.
\newblock Gamma oscillations as a biomarker for major depression: an emerging topic.
\newblock \emph{Translational Psychiatry}, 8\penalty0 (1):\penalty0 177, September 2018.
\newblock ISSN 2158-3188.
\newblock \doi{10.1038/s41398-018-0239-y}.
\newblock URL \url{https://doi.org/10.1038/s41398-018-0239-y}.

\bibitem[Trautner et~al.(2006)Trautner, Rosburg, Dietl, Fell, Korzyukov, Kurthen, Schaller, Elger, and Boutros]{trautner_sensory_2006}
Peter Trautner, Timm Rosburg, Thomas Dietl, Jürgen Fell, OA~Korzyukov, Martin Kurthen, Carlo Schaller, Christian~Erich Elger, and Nash~N Boutros.
\newblock Sensory gating of auditory evoked and induced gamma band activity in intracranial recordings.
\newblock \emph{Neuroimage}, 32\penalty0 (2):\penalty0 790--798, 2006.
\newblock \doi{10.1016/j.neuroimage.2006.04.203}.
\newblock Publisher: Elsevier.

\bibitem[Lachaux et~al.(2005)Lachaux, George, Tallon-Baudry, Martinerie, Hugueville, Minotti, Kahane, and Renault]{lachaux_many_2005}
Jean-Philippe Lachaux, Nathalie George, Catherine Tallon-Baudry, Jacques Martinerie, Laurent Hugueville, Lorella Minotti, Philippe Kahane, and Bernard Renault.
\newblock The many faces of the gamma band response to complex visual stimuli.
\newblock \emph{NeuroImage}, 25\penalty0 (2):\penalty0 491--501, 2005.
\newblock ISSN 1053-8119.
\newblock \doi{10.1016/j.neuroimage.2004.11.052}.
\newblock URL \url{https://www.sciencedirect.com/science/article/pii/S1053811904007347}.

\bibitem[Mainy et~al.(2008)Mainy, Jung, Baciu, Kahane, Schoendorff, Minotti, Hoffmann, Bertrand, and Lachaux]{mainy_cortical_2008}
Nelly Mainy, Julien Jung, Monica Baciu, Philippe Kahane, Benjamin Schoendorff, Lorella Minotti, Dominique Hoffmann, Olivier Bertrand, and Jean-Philippe Lachaux.
\newblock Cortical dynamics of word recognition.
\newblock \emph{Human brain mapping}, 29\penalty0 (11):\penalty0 1215--1230, 2008.
\newblock \doi{10.1002/hbm.20457}.
\newblock Publisher: Wiley Online Library.

\bibitem[Slutzky et~al.(2010)Slutzky, Jordan, Krieg, Chen, Mogul, and Miller]{slutzky_optimal_2010}
Marc~W Slutzky, Luke~R Jordan, Todd Krieg, Ming Chen, David~J Mogul, and Lee~E Miller.
\newblock Optimal spacing of surface electrode arrays for brain–machine interface applications.
\newblock \emph{Journal of Neural Engineering}, 7\penalty0 (2):\penalty0 026004, April 2010.
\newblock ISSN 1741-2560, 1741-2552.
\newblock \doi{10.1088/1741-2560/7/2/026004}.
\newblock URL \url{https://iopscience.iop.org/article/10.1088/1741-2560/7/2/026004}.

\bibitem[Robert~Spitzer et~al.(1989)Robert~Spitzer, Cohen, Fabrikant, and Hallett]{robert_spitzer_method_1989}
A.~Robert~Spitzer, Leonardo~G. Cohen, Judy Fabrikant, and Mark Hallett.
\newblock A method for determining optimal interelectrode spacing for cerebral topographic mapping.
\newblock \emph{Electroencephalography and Clinical Neurophysiology}, 72\penalty0 (4):\penalty0 355--361, April 1989.
\newblock ISSN 00134694.
\newblock \doi{10.1016/0013-4694(89)90072-2}.
\newblock URL \url{https://linkinghub.elsevier.com/retrieve/pii/0013469489900722}.

\bibitem[Srinivasan et~al.(1998)Srinivasan, Nunez, and Silberstein]{srinivasan_spatial_1998}
R.~Srinivasan, P.L. Nunez, and R.B. Silberstein.
\newblock Spatial filtering and neocortical dynamics: estimates of {EEG} coherence.
\newblock \emph{IEEE Transactions on Biomedical Engineering}, 45\penalty0 (7):\penalty0 814--826, July 1998.
\newblock ISSN 00189294.
\newblock \doi{10.1109/10.686789}.
\newblock URL \url{http://ieeexplore.ieee.org/document/686789/}.

\bibitem[Freeman et~al.(2003)Freeman, Holmes, Burke, and Vanhatalo]{freeman_spatial_2003}
Walter~J. Freeman, Mark~D. Holmes, Brian~C. Burke, and Sampsa Vanhatalo.
\newblock Spatial spectra of scalp {EEG} and {EMG} from awake humans.
\newblock \emph{Clinical Neurophysiology}, 114\penalty0 (6):\penalty0 1053--1068, June 2003.
\newblock ISSN 13882457.
\newblock \doi{10.1016/S1388-2457(03)00045-2}.
\newblock URL \url{https://linkinghub.elsevier.com/retrieve/pii/S1388245703000452}.

\bibitem[John et~al.(2018)John, Opie, Wong, Rind, Ronayne, Gerboni, Bauquier, O’Brien, May, Grayden, and Oxley]{john_signal_2018}
Sam~E. John, Nicholas~L. Opie, Yan~T. Wong, Gil~S. Rind, Stephen~M. Ronayne, Giulia Gerboni, Sebastien~H. Bauquier, Terence~J. O’Brien, Clive~N. May, David~B. Grayden, and Thomas~J. Oxley.
\newblock Signal quality of simultaneously recorded endovascular, subdural and epidural signals are comparable.
\newblock \emph{Scientific Reports}, 8\penalty0 (1):\penalty0 8427, 2018.
\newblock ISSN 2045-2322.
\newblock \doi{10.1038/s41598-018-26457-7}.
\newblock URL \url{https://doi.org/10.1038/s41598-018-26457-7}.
\newblock Type: Journal Article.

\bibitem[Freeman et~al.(2000)Freeman, Rogers, Holmes, and Silbergeld]{freeman_spatial_2000}
Walter~J Freeman, Linda~J Rogers, Mark~D Holmes, and Daniel~L Silbergeld.
\newblock Spatial spectral analysis of human electrocorticograms including the alpha and gamma bands.
\newblock \emph{Journal of Neuroscience Methods}, 95\penalty0 (2):\penalty0 111--121, February 2000.
\newblock ISSN 01650270.
\newblock \doi{10.1016/S0165-0270(99)00160-0}.
\newblock URL \url{https://linkinghub.elsevier.com/retrieve/pii/S0165027099001600}.

\bibitem[Nuwer(1991)]{schramm_localization_1991}
M.~R. Nuwer.
\newblock Localization of {Motor} {Cortex} with {Median} {Nerve} {Somatosensory} {Evoked} {Potentials}.
\newblock In Johannes Schramm and Aage~R. Møller, editors, \emph{Intraoperative {Neurophysiologic} {Monitoring} in {Neurosurgery}}, pages 63--71. Springer Berlin Heidelberg, Berlin, Heidelberg, 1991.
\newblock ISBN 9783642757525 9783642757501.
\newblock \doi{10.1007/978-3-642-75750-1_8}.
\newblock URL \url{http://link.springer.com/10.1007/978-3-642-75750-1_8}.

\bibitem[Schreiner et~al.(2024)Schreiner, Jordan, Sieghartsleitner, Kapeller, Pretl, Kamada, Asman, Ince, Miller, and Guger]{schreiner_mapping_2024}
Leonhard Schreiner, Michael Jordan, Sebastian Sieghartsleitner, Christoph Kapeller, Harald Pretl, Kyousuke Kamada, Priscella Asman, Nuri~F. Ince, Kai~J. Miller, and Christoph Guger.
\newblock Mapping of the central sulcus using non-invasive ultra-high-density brain recordings.
\newblock \emph{Scientific Reports}, 14\penalty0 (1):\penalty0 6527, March 2024.
\newblock ISSN 2045-2322.
\newblock \doi{10.1038/s41598-024-57167-y}.
\newblock URL \url{https://www.nature.com/articles/s41598-024-57167-y}.

\bibitem[Forsyth et~al.(2019)Forsyth, Dunston, Lombardi, Rind, Ronayne, Wong, May, Grayden, Oxley, Opie, and John]{forsyth_evaluation_2019}
Ian~A. Forsyth, Megan Dunston, Gabriel Lombardi, Gil~S. Rind, Stephen Ronayne, Yan~T. Wong, Clive~N May, David~B. Grayden, Thomas Oxley, Nicholas Opie, and Sam~E. John.
\newblock Evaluation of a minimally invasive endovascular neural interface for decoding motor activity.
\newblock In \emph{2019 9th {International} {IEEE}/{EMBS} {Conference} on {Neural} {Engineering} ({NER})}, pages 750--753, 2019.
\newblock \doi{10.1109/NER.2019.8717000}.

\bibitem[John et~al.(2017)John, Lovell, Opie, Wilson, Scordas, Wong, Rind, Ronayne, Bauquier, May, Grayden, O’Brien, and Oxley]{john_ovine_2017}
Sam~E. John, Timothy~J.H. Lovell, Nicholas~L. Opie, Stefan Wilson, Theodore~C. Scordas, Yan~T. Wong, Gil~S. Rind, Stephen Ronayne, Sébastien~H. Bauquier, Clive~N. May, David~B. Grayden, Terence~J. O’Brien, and Thomas~J. Oxley.
\newblock The ovine motor cortex: {A} review of functional mapping and cytoarchitecture.
\newblock \emph{Neuroscience \& Biobehavioral Reviews}, 80:\penalty0 306--315, September 2017.
\newblock ISSN 01497634.
\newblock \doi{10.1016/j.neubiorev.2017.06.002}.
\newblock URL \url{https://linkinghub.elsevier.com/retrieve/pii/S0149763417300726}.

\bibitem[Wolpaw and Wolpaw(2012{\natexlab{a}})]{wolpaw_braincomputer_2012-2}
Jonathan Wolpaw and Elizabeth~Winter Wolpaw.
\newblock \emph{Brain–{Computer} {Interfaces}: {Principles} and {Practice}}.
\newblock Oxford University Press, January 2012{\natexlab{a}}.
\newblock ISBN 978-0-19-538885-5.
\newblock \doi{10.1093/acprof:oso/9780195388855.001.0001}.
\newblock URL \url{http://www.oxfordscholarship.com/view/10.1093/acprof:oso/9780195388855.001.0001/acprof-9780195388855}.

\bibitem[Opie et~al.(2018)Opie, John, Rind, Ronayne, Wong, Gerboni, Yoo, Lovell, Scordas, Wilson, Dornom, Vale, O’Brien, Grayden, May, and Oxley]{opie_focal_2018}
Nicholas~L. Opie, Sam~E. John, Gil~S. Rind, Stephen~M. Ronayne, Yan~T. Wong, Giulia Gerboni, Peter~E. Yoo, Timothy J.~H. Lovell, Theodore C.~M. Scordas, Stefan~L. Wilson, Anthony Dornom, Thomas Vale, Terence~J. O’Brien, David~B. Grayden, Clive~N. May, and Thomas~J. Oxley.
\newblock Focal stimulation of the sheep motor cortex with a chronically implanted minimally invasive electrode array mounted on an endovascular stent.
\newblock \emph{Nature Biomedical Engineering}, 2\penalty0 (12):\penalty0 907--914, December 2018.
\newblock ISSN 2157-846X.
\newblock \doi{10.1038/s41551-018-0321-z}.
\newblock URL \url{https://www.nature.com/articles/s41551-018-0321-z}.

\bibitem[Oxley et~al.(2016)Oxley, Opie, John, Rind, Ronayne, Wheeler, Judy, McDonald, Dornom, Lovell, Steward, Garrett, Moffat, Lui, Yassi, Campbell, Wong, Fox, Nurse, Bennett, Bauquier, Liyanage, van~der Nagel, Perucca, Ahnood, Gill, Yan, Churilov, French, Desmond, Horne, Kiers, Prawer, Davis, Burkitt, Mitchell, Grayden, May, and O'Brien]{oxley_minimally_2016}
Thomas~J. Oxley, Nicholas~L. Opie, Sam~E. John, Gil~S. Rind, Stephen~M. Ronayne, Tracey~L. Wheeler, Jack~W. Judy, Alan~J. McDonald, Anthony Dornom, Timothy J.~H. Lovell, Christopher Steward, David~J. Garrett, Bradford~A. Moffat, Elaine~H. Lui, Nawaf Yassi, Bruce C.~V. Campbell, Yan~T. Wong, Kate~E. Fox, Ewan~S. Nurse, Iwan~E. Bennett, Sébastien~H. Bauquier, Kishan~A. Liyanage, Nicole~R. van~der Nagel, Piero Perucca, Arman Ahnood, Katherine~P. Gill, Bernard Yan, Leonid Churilov, Christopher~R. French, Patricia~M. Desmond, Malcolm~K. Horne, Lynette Kiers, Steven Prawer, Stephen~M. Davis, Anthony~N. Burkitt, Peter~J. Mitchell, David~B. Grayden, Clive~N. May, and Terence~J. O'Brien.
\newblock Minimally invasive endovascular stent-electrode array for high-fidelity, chronic recordings of cortical neural activity.
\newblock \emph{Nature Biotechnology}, 34\penalty0 (3):\penalty0 320--327, 2016.
\newblock ISSN 1546-1696.
\newblock \doi{10.1038/nbt.3428}.
\newblock URL \url{https://doi.org/10.1038/nbt.3428}.
\newblock Type: Journal Article.

\bibitem[Ross(2014)]{ross_mutual_2014}
Brian~C. Ross.
\newblock Mutual {Information} between {Discrete} and {Continuous} {Data} {Sets}.
\newblock \emph{PLoS ONE}, 9\penalty0 (2):\penalty0 e87357, February 2014.
\newblock ISSN 1932-6203.
\newblock \doi{10.1371/journal.pone.0087357}.
\newblock URL \url{https://dx.plos.org/10.1371/journal.pone.0087357}.

\bibitem[O'Toole(2020)]{otoole_mutual_2020}
John O'Toole, M.
\newblock Mutual {Information}, 2020.
\newblock URL \url{https://github.com/otoolej/mutual_info_kNN}.

\bibitem[Gerboni et~al.(2018)Gerboni, John, Rind, Ronayne, May, Oxley, Grayden, Opie, and Wong]{gerboni_visual_2018}
G~Gerboni, S~E John, G~S Rind, S~M Ronayne, C~N May, T~J Oxley, D~B Grayden, N~L Opie, and Y~T Wong.
\newblock Visual evoked potentials determine chronic signal quality in a stent-electrode endovascular neural interface.
\newblock \emph{Biomedical Physics \& Engineering Express}, 4\penalty0 (5):\penalty0 055018, August 2018.
\newblock ISSN 2057-1976.
\newblock \doi{10.1088/2057-1976/aad714}.
\newblock URL \url{https://iopscience.iop.org/article/10.1088/2057-1976/aad714}.

\bibitem[Vakani and Nair(2019)]{vakani_electrocorticography_2019}
Ravi Vakani and Dileep~R. Nair.
\newblock Electrocorticography and functional mapping.
\newblock In \emph{Handbook of {Clinical} {Neurology}}, volume 160, pages 313--327. Elsevier, 2019.
\newblock ISBN 9780444640321.
\newblock \doi{10.1016/B978-0-444-64032-1.00020-5}.
\newblock URL \url{https://linkinghub.elsevier.com/retrieve/pii/B9780444640321000205}.

\bibitem[Schalk and Leuthardt(2011)]{schalk_brain-computer_2011}
Gerwin Schalk and Eric~C. Leuthardt.
\newblock Brain-{Computer} {Interfaces} {Using} {Electrocorticographic} {Signals}.
\newblock \emph{IEEE Reviews in Biomedical Engineering}, 4:\penalty0 140--154, 2011.
\newblock ISSN 1937-3333, 1941-1189.
\newblock \doi{10.1109/RBME.2011.2172408}.
\newblock URL \url{http://ieeexplore.ieee.org/document/6047564/}.

\bibitem[Wolpaw and Wolpaw(2012{\natexlab{b}})]{wolpaw_braincomputer_2012-1}
Jonathan Wolpaw and Elizabeth~Winter Wolpaw.
\newblock \emph{Brain–{Computer} {Interfaces}: {Principles} and {Practice}, {Chapter} 7}.
\newblock Oxford University Press, January 2012{\natexlab{b}}.
\newblock ISBN 978-0-19-538885-5.
\newblock \doi{10.1093/acprof:oso/9780195388855.001.0001}.
\newblock URL \url{http://www.oxfordscholarship.com/view/10.1093/acprof:oso/9780195388855.001.0001/acprof-9780195388855}.

\bibitem[Pfurtscheller(2000)]{pfurtscheller_chapter_2000}
G.~Pfurtscheller.
\newblock Chapter 26 {Spatiotemporal} {ERD}/{ERS} patterns during voluntary movement and motor imagery.
\newblock In \emph{Supplements to {Clinical} {Neurophysiology}}, volume~53, pages 196--198. Elsevier, 2000.
\newblock ISBN 978-0-444-50499-9.
\newblock \doi{10.1016/S1567-424X(09)70157-6}.
\newblock URL \url{https://linkinghub.elsevier.com/retrieve/pii/S1567424X09701576}.

\bibitem[Mitchell et~al.(2023)Mitchell, Lee, Yoo, Morokoff, Sharma, Williams, MacIsaac, Howard, Irving, Vrljic, Williams, Bush, Balabanski, Drummond, Desmond, Weber, Denison, Mathers, O’Brien, Mocco, Grayden, Liebeskind, Opie, Oxley, and Campbell]{mitchell_assessment_2023}
Peter Mitchell, Sarah C.~M. Lee, Peter~E. Yoo, Andrew Morokoff, Rahul~P. Sharma, Daryl~L. Williams, Christopher MacIsaac, Mark~E. Howard, Lou Irving, Ivan Vrljic, Cameron Williams, Steven Bush, Anna~H. Balabanski, Katharine~J. Drummond, Patricia Desmond, Douglas Weber, Timothy Denison, Susan Mathers, Terence~J. O’Brien, J.~Mocco, David~B. Grayden, David~S. Liebeskind, Nicholas~L. Opie, Thomas~J. Oxley, and Bruce C.~V. Campbell.
\newblock Assessment of {Safety} of a {Fully} {Implanted} {Endovascular} {Brain}-{Computer} {Interface} for {Severe} {Paralysis} in 4 {Patients}: {The} {Stentrode} {With} {Thought}-{Controlled} {Digital} {Switch} ({SWITCH}) {Study}.
\newblock \emph{JAMA Neurology}, 80\penalty0 (3):\penalty0 270--278, March 2023.
\newblock ISSN 2168-6149.
\newblock \doi{10.1001/jamaneurol.2022.4847}.
\newblock URL \url{https://doi.org/10.1001/jamaneurol.2022.4847}.

\bibitem[Yanagisawa et~al.(2012)Yanagisawa, Hirata, Saitoh, Kishima, Matsushita, Goto, Fukuma, Yokoi, Kamitani, and Yoshimine]{yanagisawa_electrocorticographic_2012}
Takufumi Yanagisawa, Masayuki Hirata, Youichi Saitoh, Haruhiko Kishima, Kojiro Matsushita, Tetsu Goto, Ryohei Fukuma, Hiroshi Yokoi, Yukiyasu Kamitani, and Toshiki Yoshimine.
\newblock Electrocorticographic control of a prosthetic arm in paralyzed patients.
\newblock \emph{Annals of Neurology}, 71\penalty0 (3):\penalty0 353--361, March 2012.
\newblock ISSN 0364-5134, 1531-8249.
\newblock \doi{10.1002/ana.22613}.
\newblock URL \url{https://onlinelibrary.wiley.com/doi/10.1002/ana.22613}.

\bibitem[Yao et~al.(2022)Yao, Zhu, and Shoaran]{yao_fast_2022}
Lin Yao, Bingzhao Zhu, and Mahsa Shoaran.
\newblock Fast and accurate decoding of finger movements from {ECoG} through {Riemannian} features and modern machine learning techniques.
\newblock \emph{Journal of Neural Engineering}, 19\penalty0 (1):\penalty0 016037, February 2022.
\newblock ISSN 1741-2560, 1741-2552.
\newblock \doi{10.1088/1741-2552/ac4ed1}.
\newblock URL \url{https://iopscience.iop.org/article/10.1088/1741-2552/ac4ed1}.

\bibitem[Liao et~al.(2014)Liao, Xiao, Gonzalez, and Ding]{liao_decoding_2014}
Ke~Liao, Ran Xiao, Jania Gonzalez, and Lei Ding.
\newblock Decoding {Individual} {Finger} {Movements} from {One} {Hand} {Using} {Human} {EEG} {Signals}.
\newblock \emph{PLoS ONE}, 9\penalty0 (1):\penalty0 e85192, January 2014.
\newblock ISSN 1932-6203.
\newblock \doi{10.1371/journal.pone.0085192}.
\newblock URL \url{https://dx.plos.org/10.1371/journal.pone.0085192}.

\bibitem[Moctezuma et~al.(2019)Moctezuma, Torres-García, Villaseñor-Pineda, and Carrillo]{moctezuma_subjects_2019}
Luis~Alfredo Moctezuma, Alejandro~A. Torres-García, Luis Villaseñor-Pineda, and Maya Carrillo.
\newblock Subjects identification using {EEG}-recorded imagined speech.
\newblock \emph{Expert Systems with Applications}, 118:\penalty0 201--208, March 2019.
\newblock ISSN 09574174.
\newblock \doi{10.1016/j.eswa.2018.10.004}.
\newblock URL \url{https://linkinghub.elsevier.com/retrieve/pii/S0957417418306468}.

\bibitem[Moctezuma and Molinas(2018)]{wang_eeg-based_2018}
Luis~Alfredo Moctezuma and Marta Molinas.
\newblock {EEG}-{Based} {Subjects} {Identification} {Based} on {Biometrics} of {Imagined} {Speech} {Using} {EMD}.
\newblock In Shouyi Wang, Vicky Yamamoto, Jianzhong Su, Yang Yang, Erick Jones, Leon Iasemidis, and Tom Mitchell, editors, \emph{Brain {Informatics}}, volume 11309, pages 458--467. Springer International Publishing, Cham, 2018.
\newblock ISBN 9783030055868 9783030055875.
\newblock \doi{10.1007/978-3-030-05587-5_43}.
\newblock URL \url{http://link.springer.com/10.1007/978-3-030-05587-5_43}.

\bibitem[Lopez-Bernal et~al.(2022)Lopez-Bernal, Balderas, Ponce, and Molina]{lopez-bernal_state---art_2022}
Diego Lopez-Bernal, David Balderas, Pedro Ponce, and Arturo Molina.
\newblock A {State}-of-the-{Art} {Review} of {EEG}-{Based} {Imagined} {Speech} {Decoding}.
\newblock \emph{Frontiers in Human Neuroscience}, 16:\penalty0 867281, April 2022.
\newblock ISSN 1662-5161.
\newblock \doi{10.3389/fnhum.2022.867281}.
\newblock URL \url{https://www.frontiersin.org/articles/10.3389/fnhum.2022.867281/full}.

\bibitem[Thielen et~al.(2023)Thielen, Xu, Fujii, Rangwala, Jiang, Lin, Kammen, Liu, Selvan, Song, Mack, and Meng]{thielen_making_2023}
Brianna Thielen, Huijing Xu, Tatsuhiro Fujii, Shivani~D Rangwala, Wenxuan Jiang, Michelle Lin, Alexandra Kammen, Charles Liu, Pradeep Selvan, Dong Song, William~J Mack, and Ellis Meng.
\newblock Making a case for endovascular approaches for neural recording and stimulation.
\newblock \emph{Journal of Neural Engineering}, 20\penalty0 (1):\penalty0 011001, February 2023.
\newblock ISSN 1741-2560, 1741-2552.
\newblock \doi{10.1088/1741-2552/acb086}.
\newblock URL \url{https://iopscience.iop.org/article/10.1088/1741-2552/acb086}.

\bibitem[Kacker et~al.(2025)Kacker, Chetty, Feldman, Bennett, Yoo, Fry, Lacomis, Harel, Nogueira, Majidi, Opie, Collinger, Oxley, Putrino, and Weber]{kacker_motor_2025}
Kriti Kacker, Nikole Chetty, Ariel~K Feldman, James Bennett, Peter~E Yoo, Adam Fry, David Lacomis, Noam~Y Harel, Raul~G Nogueira, Shahram Majidi, Nicholas~L Opie, Jennifer~L Collinger, Thomas~J Oxley, David~F Putrino, and Douglas~J Weber.
\newblock Motor activity in gamma and high gamma bands recorded with a {Stentrode} from the human motor cortex in two people with {ALS}.
\newblock \emph{Journal of Neural Engineering}, 22\penalty0 (2):\penalty0 026036, April 2025.
\newblock ISSN 1741-2560, 1741-2552.
\newblock \doi{10.1088/1741-2552/adbd78}.
\newblock URL \url{https://iopscience.iop.org/article/10.1088/1741-2552/adbd78}.

\bibitem[Wolpaw and McFarland(2004)]{wolpaw_control_2004}
Jonathan~R. Wolpaw and Dennis~J. McFarland.
\newblock Control of a two-dimensional movement signal by a noninvasive brain-computer interface in humans.
\newblock \emph{Proceedings of the National Academy of Sciences}, 101\penalty0 (51):\penalty0 17849--17854, December 2004.
\newblock ISSN 0027-8424, 1091-6490.
\newblock \doi{10.1073/pnas.0403504101}.
\newblock URL \url{https://pnas.org/doi/full/10.1073/pnas.0403504101}.

\bibitem[McFarland and Wolpaw(2017)]{mcfarland_eeg-based_2017}
D.~J. McFarland and J.~R. Wolpaw.
\newblock {EEG}-based brain–computer interfaces.
\newblock \emph{Current Opinion in Biomedical Engineering}, 4:\penalty0 194--200, 2017.
\newblock ISSN 2468-4511.
\newblock \doi{10.1016/j.cobme.2017.11.004}.
\newblock Type: Journal Article.

\bibitem[Lee et~al.(2022)Lee, Schreiner, Jo, Sieghartsleitner, Jordan, Pretl, Guger, and Park]{lee_individual_2022}
Hyemin~S. Lee, Leonhard Schreiner, Seong-Hyeon Jo, Sebastian Sieghartsleitner, Michael Jordan, Harald Pretl, Christoph Guger, and Hyung-Soon Park.
\newblock Individual finger movement decoding using a novel ultra-high-density electroencephalography-based brain-computer interface system.
\newblock \emph{Frontiers in Neuroscience}, 16:\penalty0 1009878, October 2022.
\newblock ISSN 1662-453X.
\newblock \doi{10.3389/fnins.2022.1009878}.
\newblock URL \url{https://www.frontiersin.org/articles/10.3389/fnins.2022.1009878/full}.

\bibitem[Miller et~al.(2020)Miller, Hermes, and Staff]{miller_current_2020}
Kai~J. Miller, Dora Hermes, and Nathan~P. Staff.
\newblock The current state of electrocorticography-based brain–computer interfaces.
\newblock \emph{Neurosurgical Focus}, 49\penalty0 (1):\penalty0 E2, July 2020.
\newblock ISSN 1092-0684.
\newblock \doi{10.3171/2020.4.FOCUS20185}.
\newblock URL \url{https://thejns.org/view/journals/neurosurg-focus/49/1/article-pE2.xml}.

\bibitem[Metzger et~al.(2022)Metzger, Liu, Moses, Dougherty, Seaton, Littlejohn, Chartier, Anumanchipalli, Tu-Chan, Ganguly, and Chang]{metzger_generalizable_2022}
Sean~L. Metzger, Jessie~R. Liu, David~A. Moses, Maximilian~E. Dougherty, Margaret~P. Seaton, Kaylo~T. Littlejohn, Josh Chartier, Gopala~K. Anumanchipalli, Adelyn Tu-Chan, Karunesh Ganguly, and Edward~F. Chang.
\newblock Generalizable spelling using a speech neuroprosthesis in an individual with severe limb and vocal paralysis.
\newblock \emph{Nature Communications}, 13\penalty0 (1):\penalty0 6510, November 2022.
\newblock ISSN 2041-1723.
\newblock \doi{10.1038/s41467-022-33611-3}.
\newblock URL \url{https://www.nature.com/articles/s41467-022-33611-3}.

\bibitem[Silva et~al.(2024)Silva, Littlejohn, Liu, Moses, and Chang]{silva_speech_2024}
Alexander~B. Silva, Kaylo~T. Littlejohn, Jessie~R. Liu, David~A. Moses, and Edward~F. Chang.
\newblock The speech neuroprosthesis.
\newblock \emph{Nature Reviews Neuroscience}, 25\penalty0 (7):\penalty0 473--492, July 2024.
\newblock ISSN 1471-003X, 1471-0048.
\newblock \doi{10.1038/s41583-024-00819-9}.
\newblock URL \url{https://www.nature.com/articles/s41583-024-00819-9}.

\bibitem[Mahoney(2025{\natexlab{a}})]{mahoney_sep_2025}
Timothy Mahoney.
\newblock {SEP} {Stimulus} {Data}, 2025{\natexlab{a}}.
\newblock URL \url{https://figshare.unimelb.edu.au/articles/dataset/SEP_Stimulus_Data/28836644/1}.

\bibitem[Mahoney(2025{\natexlab{b}})]{mahoney_behavioural_2025}
Timothy Mahoney.
\newblock Behavioural {Dataset}, 2025{\natexlab{b}}.
\newblock URL \url{https://figshare.unimelb.edu.au/articles/dataset/Behavioural_Dataset/28836473/1}.

\end{thebibliography}

\end{document}